%% file: IEEE_TAP.tex
\documentclass[journal]{IEEEtran}
\usepackage{xcolor}
\usepackage[pdftex]{graphicx}
\graphicspath{{../pdf/}{../jpeg/}}
\DeclareGraphicsExtensions{.pdf,.jpeg,.png}
\usepackage{graphicx}
\usepackage{hyperref}
\hypersetup{
    colorlinks=true,
    linkcolor=blue,
    filecolor=blue,      
    urlcolor=black,
    citecolor =blue,
    pdftitle={Design of Perfect anomalous reflector},
    pdfpagemode=FullScreen
    }

\usepackage{subfigure}
\usepackage[cmex10]{amsmath}
\usepackage{array}
\usepackage{textcomp}
\usepackage{gensymb}
\usepackage{multirow}
\usepackage{multicol}
\usepackage{stfloats}
\usepackage[nolist]{acronym}

\begin{document}

\title{Modeling RIS from Electromagnetic Principles to \\Communication Systems--Part II: System-Level \\Simulation, Ray Tracing, and Measurement}

\author{
Le~Hao,~\IEEEmembership{Student Member,~IEEE,}
Sravan~K.~R.~Vuyyuru,~\IEEEmembership{Member,~IEEE,}
Sergei~A.~Tretyakov,~\IEEEmembership{Fellow,~IEEE}\\
Artan Salihu,~\IEEEmembership{Student Member,~IEEE,}
Markus~Rupp,~\IEEEmembership{Fellow,~IEEE,}
and Risto~Valkonen,~\IEEEmembership{Member,~IEEE,}

\thanks{This work was supported in part by the European Union’s Horizon 2020 MSCA-ITN-METAWIRELESS project, under the Marie Skłodowska-Curie grant agreement No 956256. \textit{(Corresponding author: Le~Hao)}} 
\thanks{L. Hao, A. Salihu, and M. Rupp are with TU Wien, Gusshausstrasse 25, 1040 Vienna, Austria. A. Salihu is also with Christian Doppler Laboratory for Digital Twin assisted AI for sustainable Radio Access Networks, Austria. (e-mail: \{le.hao, artan.salihu, markus.rupp\}@tuwien.ac.at).}
\thanks{S.~K.~R. Vuyyuru is with Nokia Bell Labs, Karakaari 7, 02610 Espoo, Finland
and the Department of Electronics and Nanoengineering, School of Electrical Engineering, Aalto University, 02150 Espoo, Finland (e-mail: sravan.vuyyuru@nokia.com; sravan.vuyyuru@aalto.fi).}
\thanks{S.~A. Tretyakov is with the Department of Electronics and Nanoengineering, School of Electrical Engineering, Aalto University, 02150 Espoo, Finland (e-mail: sergei.tretyakov@aalto.fi).}
\thanks{R. Valkonen is with Nokia Bell Labs, Karakaari 7, 02610 Espoo, Finland (e-mail: risto.valkonen@nokia-bell-labs.com).}
}  

\maketitle

\begin{abstract}
In this paper, we systematically study the \ac{EM} and communication aspects of a RIS through \ac{EM} simulations, system-level and ray-tracing simulations, and finally measurements. We simulate a nearly perfect, lossless RIS, and a realistic lossy \ac{AR} in different ray tracers and analyze the large-scale fading of simple RIS-assisted links. We also compare the results with continuous and quantized unit cell reflection phases with one to four-bit resolutions. Finally, we perform over-the-air communication link measurements in an indoor setting with a manufactured sample of a wide-angle \ac{AR}.
The \ac{EM}, system-level, and ray-tracing simulation results show good agreement with the measurement results. It is proved that the introduced macroscopic model of RIS from the \ac{EM} aspects is consistent with our proposed communication models, both for an ideal RIS and a realistic \ac{AR}.
\end{abstract}


\IEEEpeerreviewmaketitle

\input{Acronyms.tex}
\section{Introduction}\label{sec:Intro}
\IEEEPARstart{R}{econfigurable} Intelligent Surfaces (RIS) have attracted considerable attention in recent years. This technique is considered as an emerging technology for the next generation of wireless communications due to its potential to improve coverage and energy efficiency in wireless networks~\cite{Smart_Radio_Environments,vuyyuru2023finite,MacroscopicARM2021}. Unlike conventional reflectors or antennas, a RIS comprises multiple unit cells capable of dynamically altering their electromagnetic properties for different incoming waves. This enables RIS to actively control, redirect, and enhance electromagnetic waves in desired directions. 

The \acf{EM} perspective of the RIS-related research has been addressed in Part~I of this paper. The communication aspects, discussed in this part, focus on path loss and channel modeling for RIS-aided wireless communications. To simulate and analyze large-scale wireless networks in realistic scenarios, simulation platforms such as ray tracing and system-level simulators serve as practical tools.

Recent studies have focused on studying path-loss modeling in RIS-assisted wireless networks~\cite{Tang2021,Tang2022,Huang2022,Vittorio2022,Vitucci2023}. In~\cite{Tang2021,Tang2022}, the authors demonstrated the scaling law governing the power reflected from a RIS is influenced by various factors, such as the RIS size and the mutual distances between the RIS and the transmitter/receiver with measurements.
The authors of~\cite{Huang2022} give an overview of RIS-based channel measurements and experiments, large-scale path loss models, and small-scale multipath fading channel models, as well as channel characterization issues of RIS-assisted wireless communication systems. In~\cite{Vittorio2022}, the authors introduce a macroscopic model for evaluating the multi-mode re-radiation and diffuse scattering from a RIS. That model can be integrated with ray-based models such as ray tracing and ray launching for realistic radio propagation simulations. In addition, the authors in~\cite{Vitucci2023} extend the model to include metasurface scattering at the beginning or at the end of the interaction chain and perform ray tracing simulations in an indoor scenario for a lossy, phase-gradient \acf{AR}. The RIS-tailored Vienna \ac{SLS}~\cite{Vienna5GSLS} with a MATLAB ray tracer interface, as well as path loss models for system-level simulations have been introduced in~\cite{Hao2022, Hao2023}. In~\cite{Sihlbom2022}, the authors evaluate the system performance of a RIS-assisted cellular network through system-level simulations, such as the outdoor and indoor coverage and ergodic rate with different-sized RISs and under different frequency bands.

Even though a broad range of RIS-related research activities have been done in recent years, there is still a lack of a systematic study of RIS from the \ac{EM} design to communication models. The above mentioned models are based on the notion of the local reflection coefficient from different points of the RIS panels, but this field model is not necessarily efficient or even electromagnetically consistent, and in practice it is not possible to independently control the response of each individual array element. In addition, there are no works on analyzing a realistic RIS in a ray tracer or in a system-level simulator. It is essential to build connections between the theory and practice, as well as between the \ac{EM} design part and the communication analysis part.   

To fill the gap, in this work, we systematically study the communication link performance of a RIS that is designed based on the \ac{EM} theory of array scattering synthesis methodology~\cite{vuyyuru}. We define the appropriate controllable parameters of RIS panels and next analyze the large-scale fading of the designed RIS through the Vienna \ac{SLS} and with \ac{EM} simulation results. Moreover, we integrate the designed RIS into a ray tracer to compare the ray tracing simulation results with the theoretical outcomes. Finally, we execute measurements using a manufactured \ac{AR} prototype and compare the experimental results with theoretical analysis and ray tracing simulations. To the best knowledge of the authors, this is the first work that systematically studies a RIS from the \ac{EM} design to the system-level and ray tracing simulations, then to model validation by prototype manufacturing and link measurements. This is also the first work on implementing a perfectly designed RIS and a realistic lossy \ac{AR} to different ray tracers with the performance verified through theory. 

The remainder of this paper is organized as follows. Section~\ref{sec:SLSsimulation} introduces two methods of large-scale fading analysis and compares the results. Section~\ref{sec:RISRayTracing} explains the RIS modeling in a ray tracer and compares the ray tracing simulation with theoretical results. In Section~\ref{sec:RISExperimental}, we present  experimental results for the manufactured panel and compare with ray tracing simulations. Finally, conclusions are drawn in Section~\ref{sec:conclusion}.

\section{Large-Scale Fading Analysis} \label{sec:SLSsimulation}

In this section, we consider and compare two theoretical models of large-scale fading in RIS-assisted links. At this stage, we assume a far-field propagation scenario with a single \ac{LOS}-path communication link. One of the studied methods is based on a theoretical estimation of the response of perfectly functioning \ac{AR}s \cite{Sergei2023}, incorporated in the Vienna \ac{SLS}, and the other one is based on numerically simulated RIS directivity patterns. 

\subsection{Method 1}\label{sec:method1}
A recently published path loss model~\cite{Sergei2023}, derived from an approximate electromagnetic solution for scattered fields from RIS, has been implemented in the Vienna \ac{SLS}. This model is designed only for far-field propagation scenarios, and it is not applicable to near-field cases. Therefore, in this paper, we only analyze the far-field performance of RISs, and the near-field analysis is postponed to our future work. With this path loss model, the received power at the RX antenna is calculated as 
\begin{equation}\label{equ:eq_linkbudget}
P_r = P_t G_t(\theta_t, \phi_t) G_r(\theta_r, \phi_r) \eta_{\text{eff}} \left(\frac{S_1}{4\pi R_1 R_2}\right)^2 |\cos\theta^i \cos\theta^r|,
\end{equation}
where $0<\eta_{\text{eff}}\leq 1$ is the RIS efficiency parameter that takes into account parasitic absorption in RIS as well as design and manufacturing imperfections. $S_1$ is the geometrical area of the RIS panel. The parameters $\theta^i$ and $\theta^r$ represent the incidence and reflection angles at the position of the RIS, respectively. 
The transmit power is indicated as $P_t$, while $G_t(\theta_t, \phi_t)$ and $G_r(\theta_r, \phi_r)$ represent the gains of the TX and RX antennas, respectively. $\theta_t$, $\phi_t$, $\theta_r$, and $\phi_r$ represent the elevation angle and the azimuth angle from the TX antenna to the RIS, and the elevation angle and the azimuth angle from the RX antenna to the RIS, respectively. The distance between the base station (TX) and the RIS is denoted by $R_1$, while the distance between the RIS and the user (RX) is denoted by $R_2$. 

\subsection{Method 2}\label{sec:method2}
Another path loss model for a RIS-assisted link is based on the notion of RIS directivity and gain. The directivity is defined in terms of the electric field far-field pattern $F(\theta, \phi)$ as~\cite{balanis2015antenna}: 
\begin{equation}\label{equ:directivity}
D(\theta, \phi) = \frac{4\pi F(\theta, \phi)}{\int_{0}^{2\pi}\int_{0}^{\pi} F(\theta, \phi) \sin\theta d\theta d\phi},
\end{equation}
where $F(\theta, \phi)$ is the far-zone radiation intensity pattern.
The gain is calculated as
\begin{equation}\label{equ:antgain}
G(\theta, \phi) = e_{\text{cd}} D(\theta, \phi),
\end{equation}
where $e_{\text{cd}}$ is the panel efficiency. If RIS losses can be neglected, we have $e_{\text{cd}} = 1$. 

In this work, we consider the designed RIS from Part~I and calculate its gain numerically, using CST software. The RIS gain values are found for four different modes each for five different sizes of RIS panels. The RIS gain results for the continuous load impedance design are listed in Table~\ref{tab:risgain}. It is worth noting that each RIS model needs two gain values: $G_\text{rx}$ is the RIS gain in the direction from RIS to TX, and $G_\text{tx}$ is the RIS gain in the direction from RIS to RX. Corresponding to our RIS designed for the normal incidence angle and four reflection angles, $G_\text{rx}$ is obtained at $0^{\degree}$, and $G_\text{tx}$ is obtained at $13^{\degree}$, $27^{\degree}$, $43^{\degree}$, and $65^{\degree}$.  

\begin{table*}[h]
\caption{Different-size RIS gains for continuous and quantized loads}
\label{tab:risgain}
\centering
\begin{tabular}{| c | c | c | c | c | c| c| c | c | c | c| c|}
\hline
\multirow{2}{4em}{Floquet mode} & \multirow{2}{4em}{Resolution} & \multicolumn{2}{c|}{$32\times32$} & \multicolumn{2}{c|}{$48\times48$} & \multicolumn{2}{c|}{$64\times64$} & \multicolumn{2}{c|}{$80\times80$} & \multicolumn{2}{c|}{$96\times96$}\\ \cline{3-12}
 & & $G_\text{tx}$~(dB) & $G_\text{rx}$~(dB) & $G_\text{tx}$~(dB) & $G_\text{rx}$~(dB) & $G_\text{tx}$~(dB) & $G_\text{rx}$~(dB) & $G_\text{tx}$~(dB) & $G_\text{rx}$~(dB) & $G_\text{tx}$~(dB) & $G_\text{rx}$~(dB) \\ \hline
 
\multirow{5}{4em}{Mode $1$ $(13^{\degree})$} & Continuous  & $29.86$ & $30.04$ & $33.36$ & $33.61$ & $35.84$ & $36.19$ & $37.76$ & $38.27$ & $39.33$ & $40.04$\\ \cline{2-12}
 &  $4$ bit  & $29.86$ & $30.03$ & $33.35$ & $33.61$ & $35.84$ & $36.19$ & $37.76$ & $38.26$ & $39.32$ & $40.03$ \\ \cline{2-12}
 &  $3$ bit  & $29.78$ & $29.95$ & $33.28$ & $33.52$ & $35.76$ & $36.10$ & $37.68$ & $38.18$ & $39.26$ & $39.95$ \\ \cline{2-12}
 &  $2$ bit  & $29.57$ & $29.73$ & $33.06$ & $33.30$ & $35.54$ & $35.88$ & $37.47$ & $37.95$ & $39.04$ & $39.70$ \\ \cline{2-12}
 &  $1$ bit  & $27.20$ & $27.28$ & $30.69$ & $30.81$ & $33.17$ & $33.34$ & $35.10$ & $35.35$ & $36.68$ & $37.02$ \\ \hline
 
\multirow{5}{4em}{Mode $2$ $(27^{\degree})$} & Continuous  & $29.49$ & $30.04$ & $32.99$ & $33.61$ & $35.47$ & $36.19$ & $37.39$ & $38.27$ & $38.97$ & $40.04$ \\ \cline{2-12}
 & $4$ bit  & $29.45$ & $30.01$ & $32.95$ & $33.58$ & $35.43$  & $36.16$ & $37.36$ & $38.23$ & $38.93$ & $40.00$\\ \cline{2-12}
 &  $3$ bit  & $29.45$ & $29.99$ & $32.94$ & $33.56$ & $35.42$ & $36.14$ & $37.35$ & $38.22$ & $38.93$ & $39.98$ \\ \cline{2-12}
 &  $2$ bit  & $29.23$ & $29.76$ & $32.71$ & $33.33$ & $35.20$ & $35.90$ & $37.12$ & $37.97$ & $38.70$ & $39.72$ \\ \cline{2-12}
 &  $1$ bit  & $26.46$ & $27.03$ & $29.94$ & $30.53$ & $32.42$ & $33.04$ & $34.35$ & $35.02$ & $35.94$ & $36.68$ \\ \hline
 
\multirow{5}{4em}{Mode $3$ $(43^{\degree})$} & Continuous  & $28.69$ & $30.03$ & $32.16$ & $33.61$ & $34.63$ & $36.19$ & $36.54$ & $38.26$ & $38.09$ & $40.03$\\ \cline{2-12}
 & $4$ bit  & $28.51$ & $29.87$ & $31.99$ & $33.44$ & $34.45$  & $36.01$ & $36.36$ & $38.09$ & $37.92$ & $39.84$\\ \cline{2-12}
 &  $3$ bit  & $28.51$ & $29.86$ & $31.98$ & $33.43$ & $34.45$ & $36.01$ & $36.36$ & $38.08$ & $37.93$ & $39.84$ \\ \cline{2-12}
 &  $2$ bit  & $28.37$ & $29.68$ & $31.82$ & $33.24$ & $34.28$ & $35.81$ & $36.19$ & $37.87$ & $37.74$ & $39.62$ \\ \cline{2-12}
 &  $1$ bit  & $24.97$ & $26.37$ & $28.46$ & $29.88$ & $30.94$ & $32.39$ & $32.87$ & $34.38$ & $34.44$ & $36.03$ \\ \hline
 
\multirow{5}{4em}{Mode $4$ $(65^{\degree})$} & Continuous  & $26.72$ & $30.04$ & $30.07$ & $33.61$ & $32.47$ & $36.19$ & $34.35$ & $38.27$ & $35.90$ & $40.04$\\ \cline{2-12}
 & $4$ bit & $26.65$ & $30.00$ & $30.01$ & $33.57$ & $32.41$  & $36.15$ & $34.29$ & $38.22$ & $35.84$ & $39.98$\\ \cline{2-12}
 &  $3$ bit  & $26.16$ & $29.56$ & $29.52$ & $33.11$ & $31.93$ & $35.67$ & $33.81$ & $37.72$ & $35.36$ & $39.46$ \\ \cline{2-12}
 &  $2$ bit  & $25.97$ & $29.36$ & $29.32$ & $32.90$ & $31.73$ & $35.45$ & $33.61$ & $37.50$ & $35.17$ & $39.22$ \\ \cline{2-12}
 &  $1$ bit  & $22.31$ & $25.89$ & $25.70$ & $29.34$ & $28.13$ & $31.81$ & $30.06$ & $33.77$ & $31.65$ & $35.40$ \\ \hline 
\end{tabular}\vspace{-1em}
\end{table*}

Once we obtain the RIS gains from CST simulations, the received power at the RX antenna through the RIS can be calculated according to Friis' formula for the links between TX and RIS and then between RIS and RX~\cite{balanis2015antenna}:
\begin{equation}\label{equ:P1}
P_1 = G_t(\theta_t, \phi_t) G_\text{rx}(\theta_{\text{rx}}, \phi_{\text{rx}}) \left(\frac{\lambda}{4\pi R_1}\right)^2,
\end{equation}
\begin{equation}\label{equ:P2}
P_2 = G_\text{tx}(\theta_{\text{tx}}, \phi_{\text{tx}}) G_r(\theta_r, \phi_r) \left(\frac{\lambda}{4\pi R_2}\right)^2,
\end{equation}
which gives the following path loss estimation~\cite{Ellingson2021}: 
\begin{align}\label{equ:pr}
    P_r &= P_t P_1 P_2 \nonumber \\
    &= \frac{ P_t G_t(\theta_t, \phi_t) G_\text{rx}(\theta_{\text{rx}}, \phi_{\text{rx}}) G_\text{tx}(\theta_{\text{tx}}, \phi_{\text{tx}}) G_r(\theta_r, \phi_r) \lambda^4}{(4\pi)^4(R_1 R_2)^2}.
\end{align}
Here, $\theta_{\text{rx}}$, $\phi_{\text{rx}}$, $\theta_{\text{tx}}$, and $ \phi_{\text{tx}}$ represent the spherical angles defined
from the RIS to the TX antenna, and from the RIS to the RX antenna, respectively.

\subsection{Comparison Between the Analytical Path Loss Model and Simulations} \label{sec:2mtdcompare}
\begin{figure}[t]
\centering		
\subfigure[\label{Fig:1a}]{\includegraphics[width=0.241\textwidth]{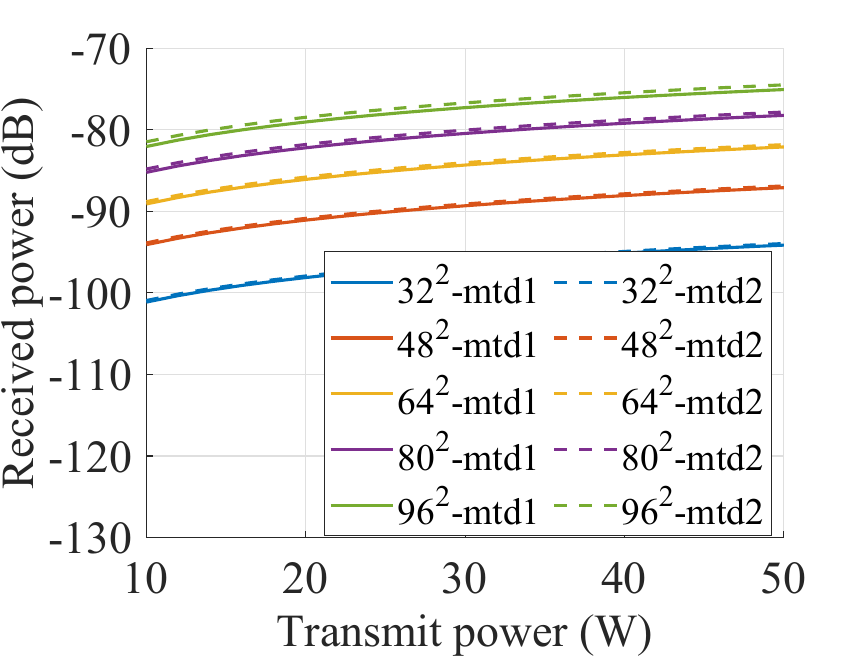}}
\subfigure[\label{Fig:1b}]{\includegraphics[width=0.241\textwidth]{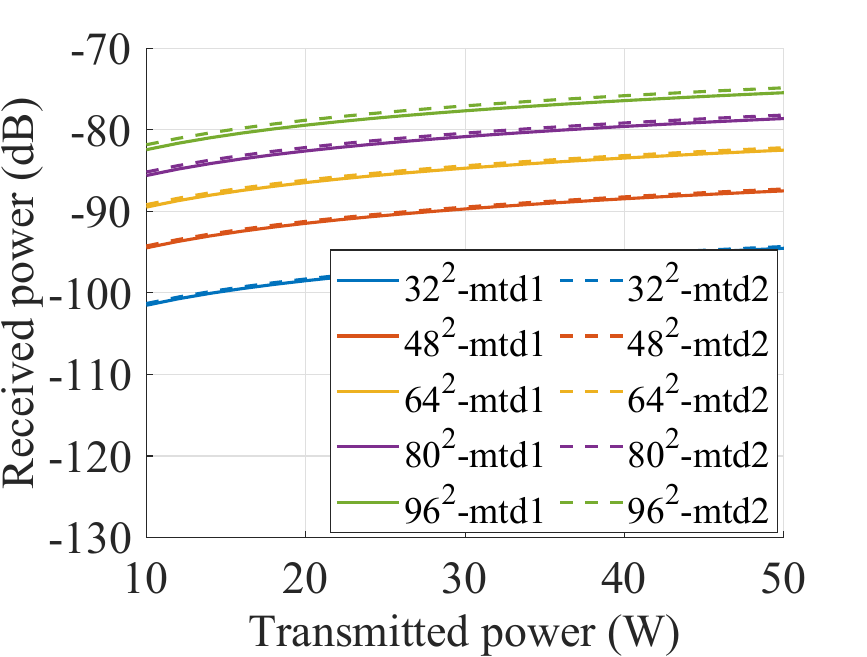}} 
\subfigure[\label{Fig:1c}]{\includegraphics[width=0.241\textwidth]
{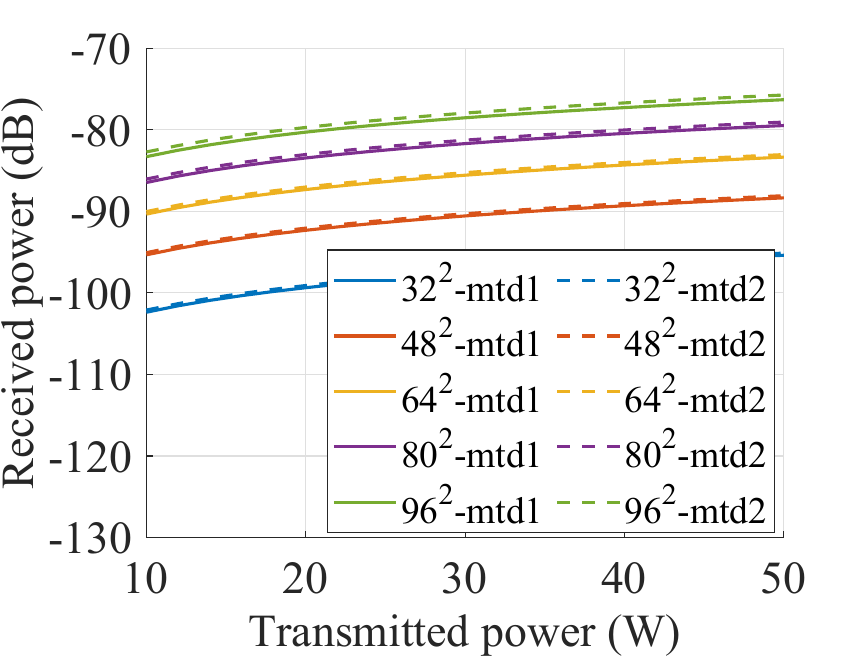}}
\subfigure[\label{Fig:1d}]{\includegraphics[width=0.241\textwidth]{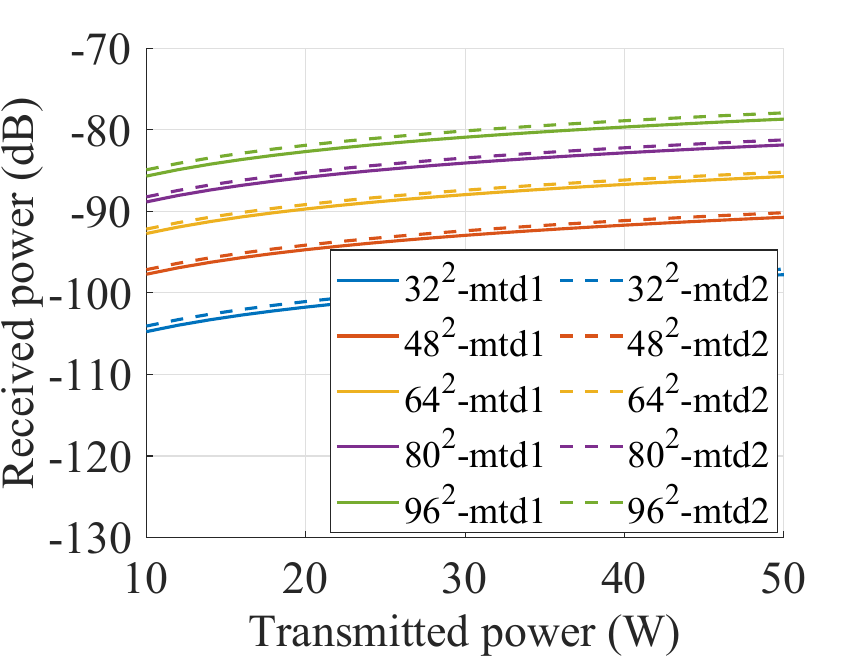}}
\caption{Results comparison between the two methods with different RIS sizes. (a) Mode $1$, (b) Mode $2$, (c) Mode $3$, (d) Mode $4$. \label{Fig:pr_2mtd}}\vspace{-1.5em}
\end{figure}

Here, we compare the path loss estimations obtained using both methods for a simple case of a single LOS link TX -- RIS -- RX. 
As an example, we set $G_t = G_r = 1$, $R_1 = 17$~m and $R_2 = 17.22$~m, $\eta_{\text{eff}} = 1$. According to the design parameters of the RIS, the working angles are $\theta^i = 0^{\degree}$ and $\theta^r = 13^{\degree}$, $27^{\degree}$, $43^{\degree}$, $65^{\degree}$ for modes $1$ to $4$, respectively. The test RIS areas are $S_1 = 32\times32~A$, $48\times48~A$, $64\times64~A$, $80\times80~A$, and $96\times96~A$ where $A=(1.1034\lambda /4)^2$ is the area of one unit cell. The common parameters in Eq.~(\ref{equ:pr}) are set the same as Eq.~(\ref{equ:eq_linkbudget}). $G_\text{rx}$ and $G_\text{tx}$ in Eq.~(\ref{equ:pr}) are the computed  values from Table~\ref{tab:risgain} with continuous loads. The results of the received power from the two methods are shown in Figs.~\ref{Fig:1a}, \ref{Fig:1b}, \ref{Fig:1c}, and \ref{Fig:1d} for modes $1$ to $4$, respectively.

From Fig.~\ref{Fig:pr_2mtd} we can observe that for all four modes, the two methods give very close results. The differences between the two methods are about $0.2$~dB to $0.6$~dB for mode $1$, mode $2$, and mode $3$ when the RIS sizes change from $32\times32$ to $96\times96$. For mode $4$ the difference is from $0.7$~dB to $0.8$~dB with the five sizes. 

This agreement is expected because Eq.~\eqref{equ:eq_linkbudget} is valid for theoretically perfect \ac{AR}s, and from Part I we saw that the RIS design with continuous loads gives a nearly perfect performance. In fact, it can be shown that for ideal \ac{AR}s with continuous current distribution the considered two path loss models are equivalent. The model of \eqref{equ:eq_linkbudget} assumes that the RIS captures all the power that is incident on its surface and retransmits it without imperfections. This means that if we consider the same RIS as a conjugate-matched receiving antenna, its effective area $A_{\text{eff}}$ is equal to the geometrical area of the panel cross-section, that is,  
$
A_{\text{eff}}=S_1|\cos\theta^i|
$. 
Likewise, in the transmit regime, we have 
$
A_{\text{eff}}=S_1|\cos\theta^r|$.
Using the general relation between the effective area and gain, valid for any linear and reciprocal antenna,
\begin{equation}\label{equ:risgain} 
G = 4\pi\frac{ A_{\text{eff}}}{\lambda^2},
\end{equation}
We can find the RIS gains for an ideal \ac{AR} in terms of the panel area and the incidence and reflection angles: 
\begin{equation}\label{equ:gainrx}
G_\text{rx} = 4\pi\frac{ S_1}{\lambda^2}|\cos\theta^i|
\end{equation}
and
\begin{equation}\label{equ:gaintx}
G_\text{tx} = 4\pi\frac{ S_1}{\lambda^2}|\cos\theta^r|.
\end{equation}
Substituting Eq.~(\ref{equ:gainrx}) and (\ref{equ:gaintx}) into Eq.~(\ref{equ:pr}) we obtain the same equation as Eq.~(\ref{equ:eq_linkbudget}). Therefore, these two methods are equivalent if the RIS operates perfectly. 
The small differences between the two methods are from the RIS gain differences between CST simulation and the ideal theoretical values given by Eqs.~(\ref{equ:gainrx}) and (\ref{equ:gaintx}), and they result from the spatial discretization of the reflecting surface.

From~\cite{Hao2023} we conclude that when the RIS size is doubled, the received power should achieve $6$~dB gain for a tuned RIS. In these four figures, the received power has about $7$~dB, $5$~dB, $4$~dB and $3$~dB differences for the RIS sizes  $32\times32$ to $48\times48$, from $48\times48$ to $64\times64$ from $64\times64$ to $80\times80$, and from $80\times80$ to $96\times96$, respectively. Since the RIS size $64\times64$ is four times larger than the size $32\times32$, the received power with $64\times64$ size is $12$~dB higher than for the $32\times32$-sized RIS. Similarly, the difference between $48\times48$ and $96\times96$-sized RIS is also $12$~dB, which is consistent with the power scaling law~\cite{Wu2018}.

\subsection{Load Quantization Analysis}\label{sec:quantization}
The RIS gain values used in Sec.~\ref{sec:2mtdcompare} are from the optimization of continuous reactive loads, corresponding to the assumption that the controllable loads can have arbitrary reactive impedances. In this section, we use the RIS gain results obtained from quantized load impedances, which are summarized in Table~\ref{tab:risgain}, to investigate the difference between different quantization resolutions. From Part I we explained that the reflection efficiency of the RIS increases when the load quantization resolution changes from $1$-bit to $4$-bit. This is because the unit cell loads optimization results become more efficient when we have more load impedance values ($2$ values for $1$-bit and $16$ values for $4$-bit). The reflection wave is more concentrated in the desired directions and the side lobes are better suppressed, which is why the RIS gain values in the desired directions also become higher and gradually get close to the gain when using continuous loads.

\begin{figure}[t]
\centering		\vspace{-1.5em}
\subfigure[\label{Fig:2a}]{\includegraphics[width=0.241\textwidth]{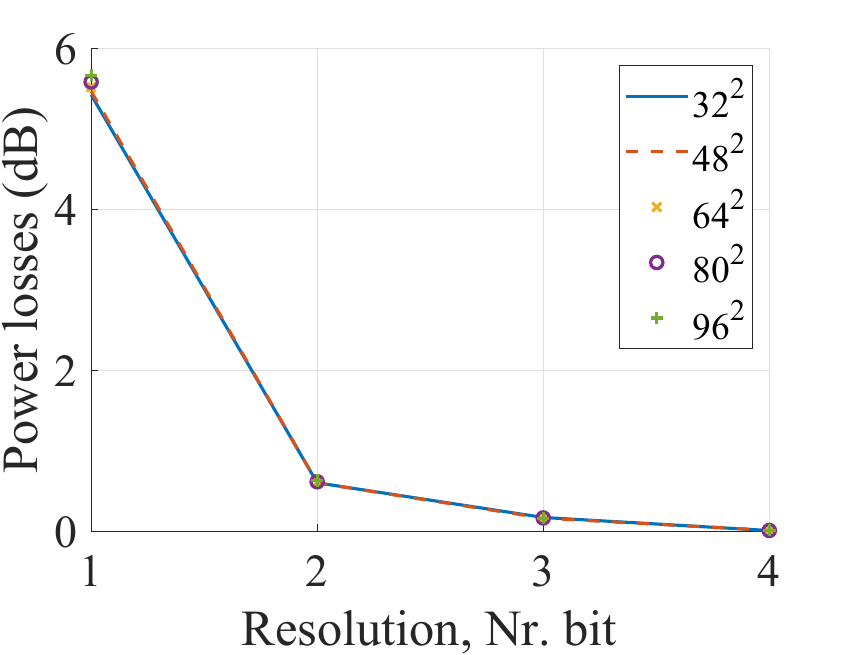}}
\subfigure[\label{Fig:2b}]{\includegraphics[width=0.241\textwidth]{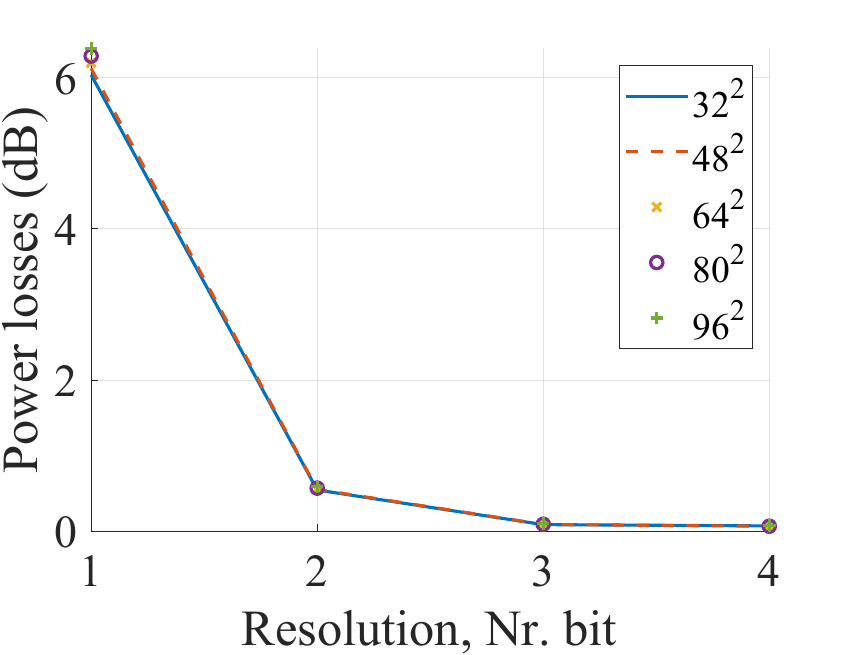}}
\subfigure[\label{Fig:2c}]{\includegraphics[width=0.241\textwidth]{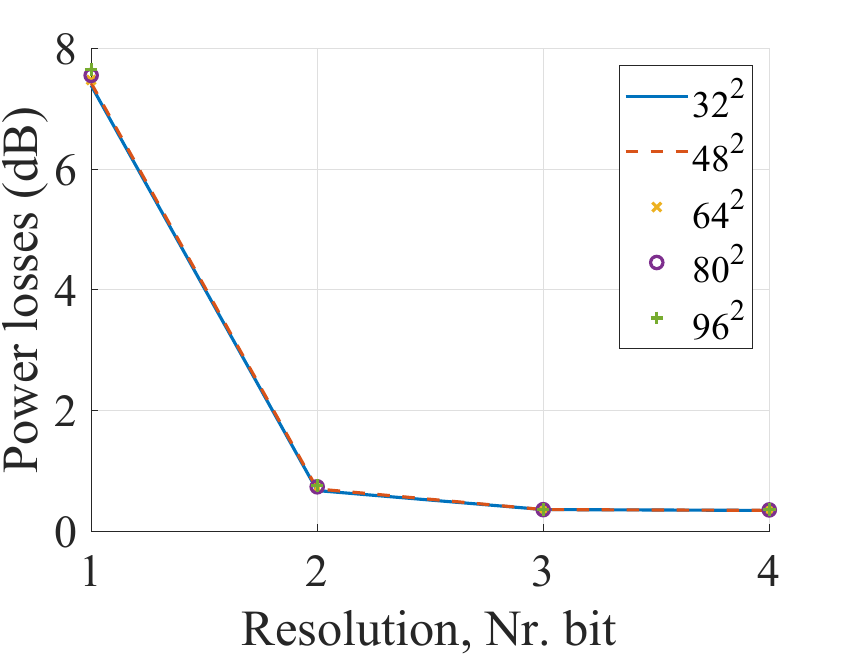}}
\subfigure[\label{Fig:2d}]{\includegraphics[width=0.241\textwidth]{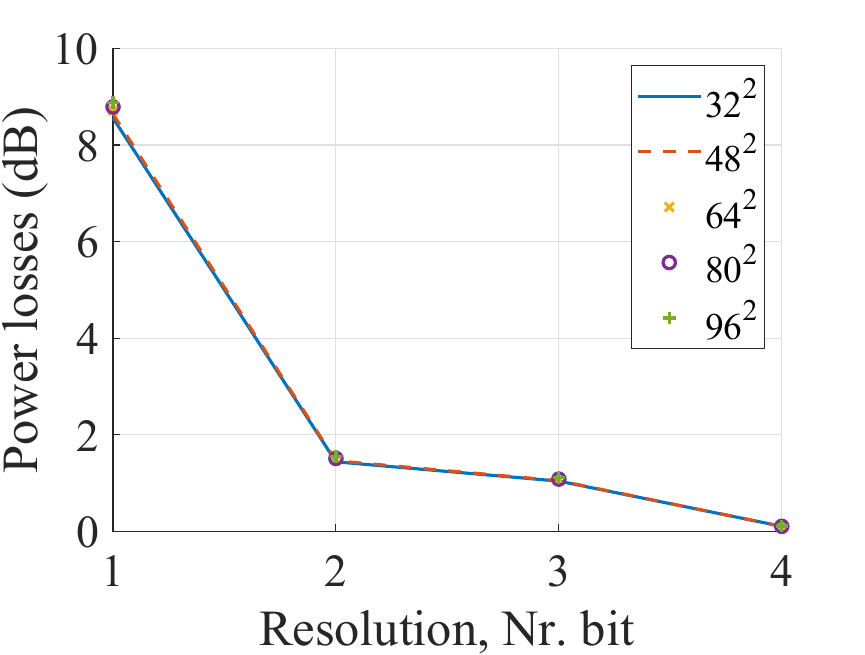}}
\caption{Power losses due to load quantization.  (a) Mode $1$, (b) Mode $2$, (c) Mode $3$, (d) Mode $4$. \label{Fig:rxpower_quantization1}}
\vspace{-1.5em}
\end{figure}

In Fig.~\ref{Fig:rxpower_quantization1} we compare the received powers between the designs based on continuous and quantized load values. Figures~\ref{Fig:2a}, \ref{Fig:2b}, \ref{Fig:2c}, and \ref{Fig:2d} show the results for modes $1,2,3,4$, respectively. The differences between the panels of $5$ different sizes are almost the same for all four modes. As expected, when the resolution increases from $1$-bit to $4$-bit, the differences between the continuous loads designs and the discrete loads become smaller for all four modes.  
It can be observed that for $1$-bit resolution, the scattering losses are quite high for all four modes, while the $4$-bit resolution leads to very similar results as for the continuous loads. The $3$-bit resolution gives already relatively good results, i.e., $1.11$~dB for mode $4$, and less than $0.4$~dB loss for modes $3,2$, and $1$ are good enough.

\section{Ray Tracing Simulations}
\label{sec:RISRayTracing}
The results in Sec.~\ref{sec:SLSsimulation} are based on the free-space path loss model, 
with only one \ac{LOS} path between the TX and RIS, and one \ac{LOS} path between the RIS and RX. 
To analyze wave propagation in more realistic environments, ray tracing is a very useful method since it accounts for the effect of the environment. There are several ray tracers already in the academic and industrial use, such as the MATLAB ray tracer, the Wireless InSite from Remcom, the CloudRT from Beijing Jiaotong University, and more. However, there are still no ray tracers that would include an accurately modeled RIS module. To simulate a RIS-assisted scenario, we have to first model a RIS into the ray tracer. In this section, we first incorporate a model of the  designed RIS in the Wireless InSite ray tracer and verify the simulated results against the theory. After that, we extend the simulation scenario from a simple \ac{SISO} case to a multipath scenario and analyze the simulation results. 

\subsection{Verification in a SISO Scenario}
\label{subsec:RIS_SISO_Scenario} 
We utilize Wireless InSite to accommodate RIS functionality by modeling the RIS as two separate antennas with imported E-field data from CST.
For each RIS size at each propagation mode, we have two RIS patterns, one towards the incidence direction and the other toward the realized anomalous reflection angle. Therefore, in the ray tracer, we first simulate the TX-RIS link where the RIS is used as a receiver. Next, we simulate the RIS-RX link where the RIS is used as a transmitter. 

\begin{figure}[t]
\center 
\includegraphics[width=0.4\textwidth]{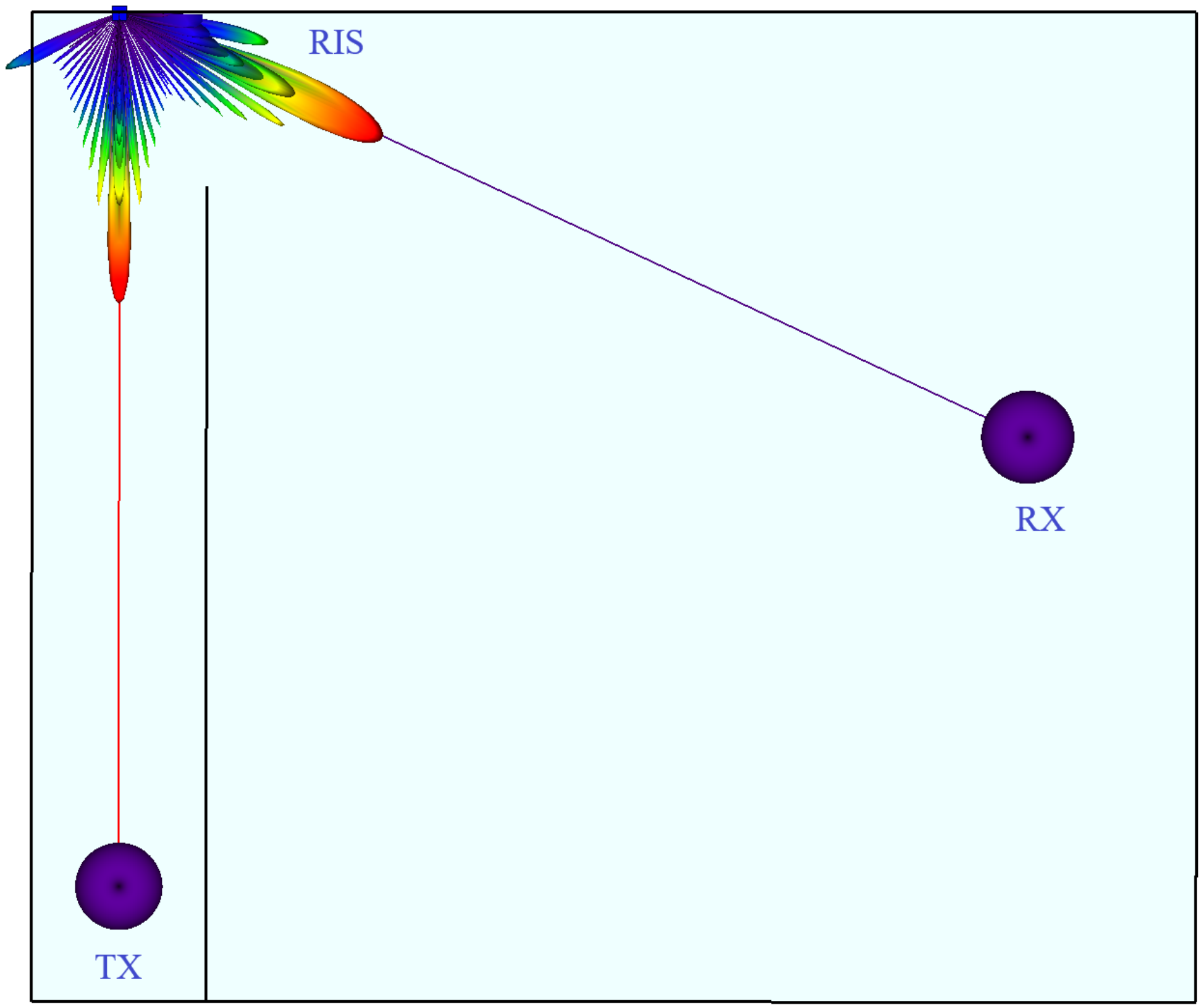}
\caption{A SISO scenario with RIS in Wireless InSite.}
\label{Fig:remcomSISOscenario} 
\end{figure}

To verify the RIS modeling in the ray tracer, we set up the same SISO scenario in Wireless InSite as in Sec.~\ref{sec:SLSsimulation}, see  Fig.~\ref{Fig:remcomSISOscenario}. The center frequency is $26$~GHz. The TX and RX antennas are initially omnidirectional ($0$~dB gain). The distance between the TX and RIS is $17$~m, and the distance between the RIS and RX is $17.22$~m. When the RIS is used as a receiver for the TX-RIS link, the RIS pattern is toward $0^\degree$, facing the TX antenna. When the RIS is used as a transmitter for the RIS-RX link, the reflection pattern of the RIS is towards $13^\degree, 27^\degree, 43^\degree,$ and $65^\degree$ for modes $1,2,3,4$, respectively. The TX and RX direct link is  blocked by a wall, so that there is no \ac{LOS} path between them. First, the reflection path number is set as $0$, so that we only observe the \ac{LOS} paths  TX-RIS and RIS-RX. All walls, ceiling and floor in this scenario are considered as perfect absorbers, to provide a direct point of comparison with the \ac{LOS} path loss models considered above.

\begin{figure}[t]
\centering		
\subfigure[\label{Fig:3a}]{\includegraphics[width=0.241\textwidth]{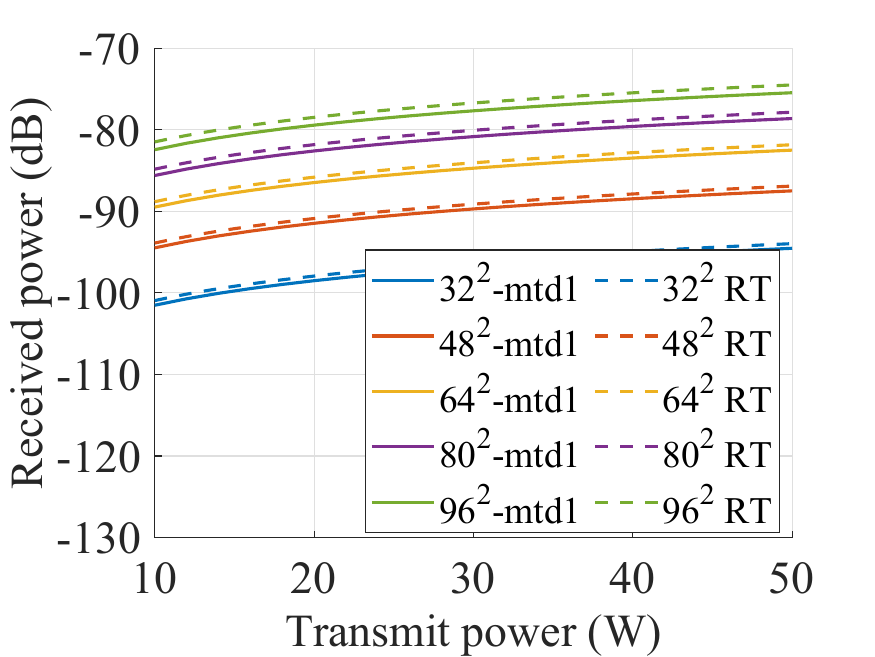}}
\subfigure[\label{Fig:3b}]{\includegraphics[width=0.241\textwidth]{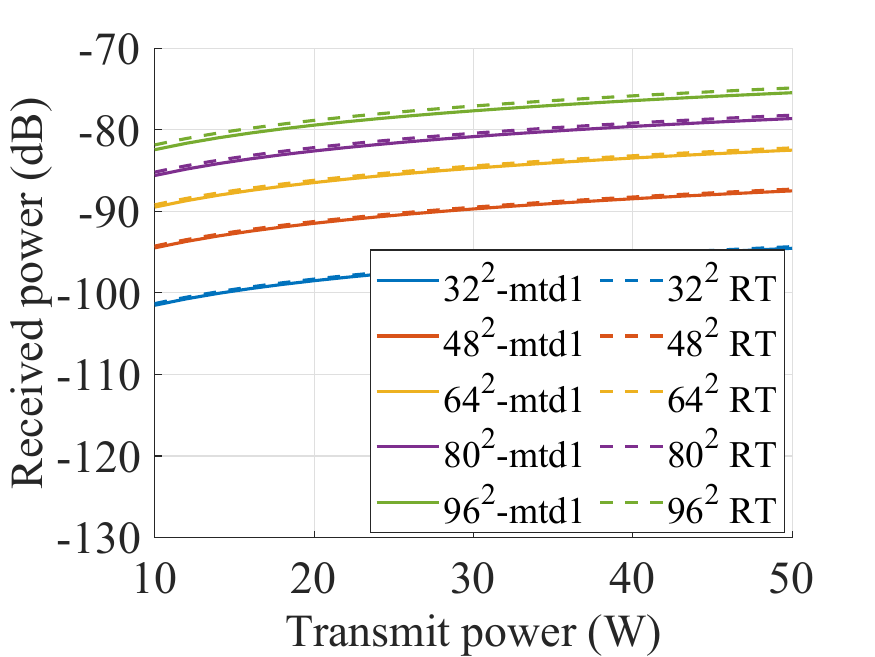}}
\subfigure[\label{Fig:3c}]{\includegraphics[width=0.241\textwidth]{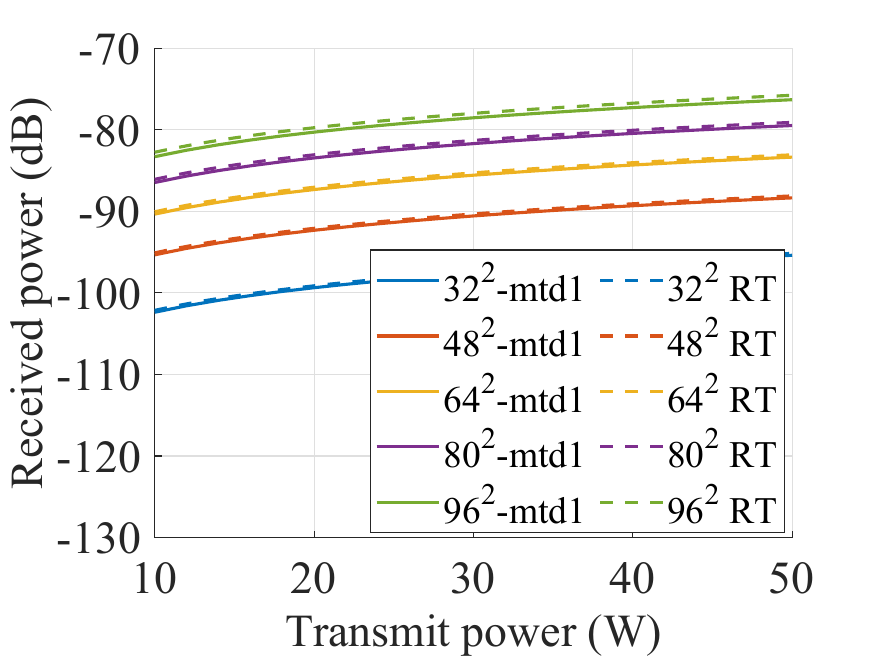}}
\subfigure[\label{Fig:3d}]{\includegraphics[width=0.241\textwidth]{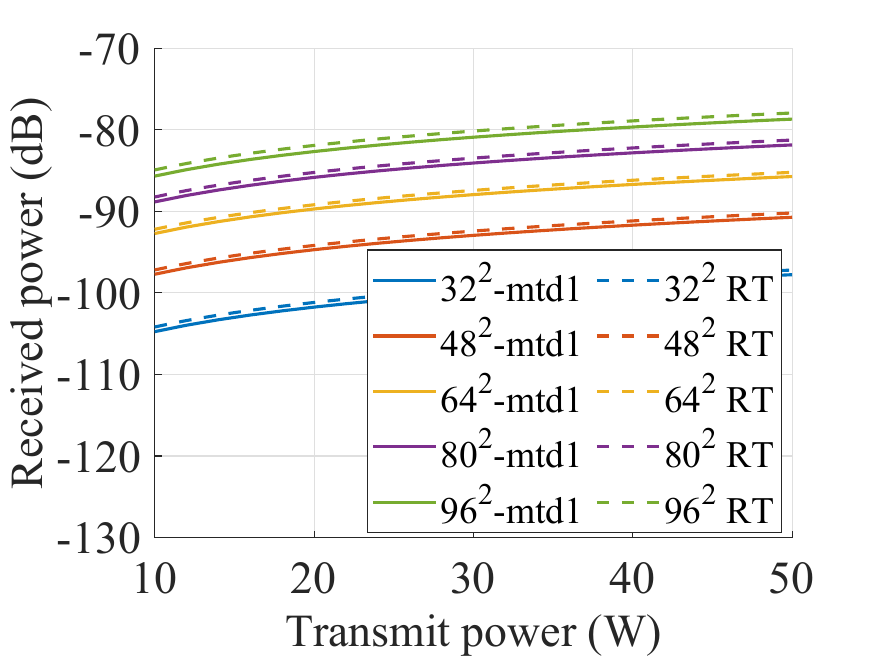}}
\caption{Results comparison between method $1$ and ray tracing for different RIS sizes. (a) Mode $1$, (b) Mode $2$, (c) Mode $3$, (d) Mode $4$. \label{Fig:com_mtd1_RT}}
\vspace{-1.5em}
\end{figure}

The comparison results between the ray tracing simulation and method $1$ from Sec.~\ref{sec:method1} are plotted in Fig.~\ref{Fig:com_mtd1_RT}. Figures~\ref{Fig:3a}, \ref{Fig:3b}, \ref{Fig:3c}, and \ref{Fig:3d} show results for modes $1,2,3,4$, respectively. From the four figures we can observe that the ray tracing simulation results are very close to the theoretical results that we have obtained for method $1$. The larger the RIS size, the larger the differences between the two results for all four modes. However, even the largest difference that appears for mode $4$ is about $0.6$~dB. The comparison results indicate that our strategy of modeling RIS in the ray tracer seems correct for the \ac{LOS} link.

\subsection{Multi-user Scenario with only LOS Paths}
\label{sec:LOSRTsimulation}
\begin{figure}[t]
\center
\includegraphics[width=0.4\textwidth]{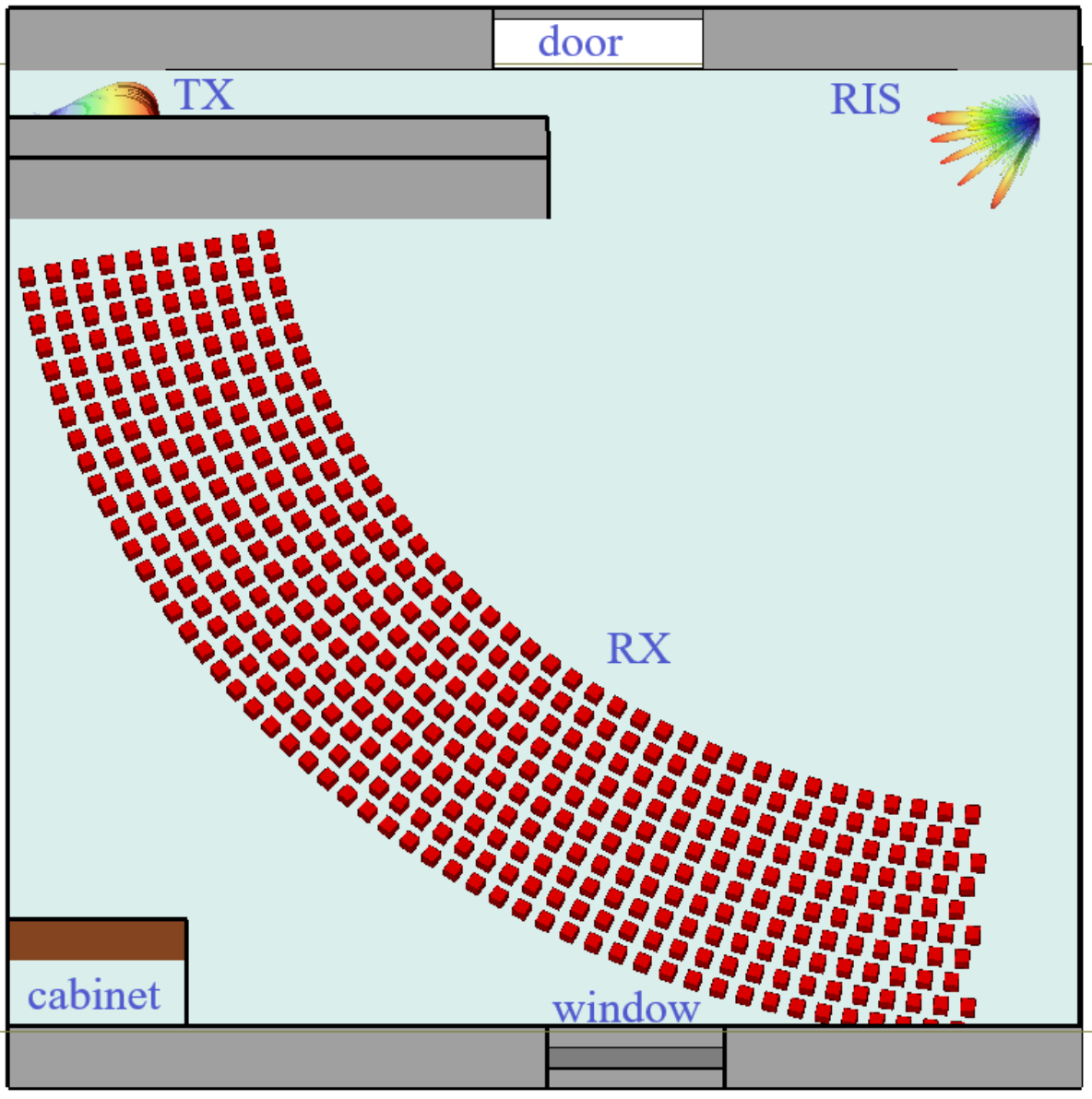}
\caption{An indoor scenario with RIS in the Wireless InSite software.}
\label{Fig:RemcomSceanrio}
\end{figure}

In this section, we extend the simulation to a multi-user scenario, as shown in Fig.~\ref{Fig:RemcomSceanrio}. The room size is $24\times25\times3$ (m$^3$) in terms of width $\times$ length $\times$ height. There is one $1\times 1$~m$^2$ glass window on the southern wall, and a $1.3\times 2.5$~m$^2$ door on the northern wall. The material of ceiling and floor is concrete, and the material of all walls is layered drywall. In the southwest corner of the room, there is a small wooden cabinet with a height of $2$~m. The direct link between the TX and RX antennas is blocked by two inner walls in the room. The TX antenna is a horn antenna with the maximum gain of $18.5$~dBi toward $0^{\degree}$ from the RIS. To investigate how the RX antenna location influences the received power, we place $450$ test omnidirectional RX antennas at different locations in the room, shown as red cubes. The spacing between the adjacent RX antennas is $0.6$~m. 
The RISs for mode $1$ to mode $4$ with five different sizes are placed at the same location in the room with their receiving beams toward the TX antenna. The far-field distances of the five RIS sizes are $1.80, 4.04, 7.19, 11.23$, and $16.17$~m for the sizes of $32\times32$, $48\times48$, $64\times64$, $80\times80$, and $96\times96$, respectively, according to the calculation $R=2D^2/\lambda$ with $D$ the largest dimension of the antenna. Hence, the distance between the TX and the RIS is set to $22$~m, and the distance between the RIS and the RX antenna is from $17.4$~m to $22.8$~m to fulfil the far-field assumption. The RX antennas are placed at $10$ arcs with the RIS location being the center point of the arcs. The angle range of the RX antennas toward the RIS is from $10^\circ$ to $85.4^\circ$ from northwest to southeast in the room. The height of TX, RX, and the RIS is $1.5$~m.

\begin{figure}[t]
\centering		
\subfigure[\label{Fig:4a}]{\includegraphics[width=0.241\textwidth]{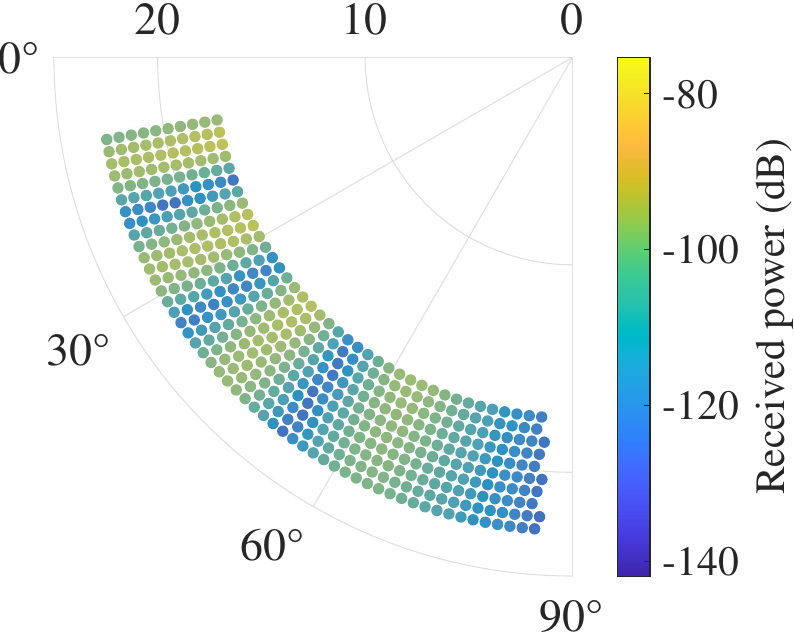}}
\subfigure[\label{Fig:4b}]{\includegraphics[width=0.241\textwidth]{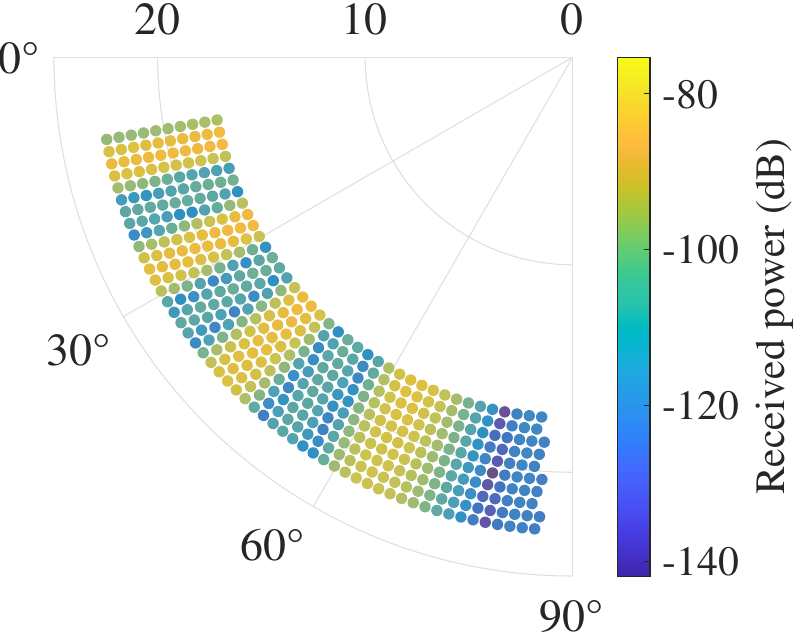}}
\subfigure[\label{Fig:4c}]{\includegraphics[width=0.241\textwidth]{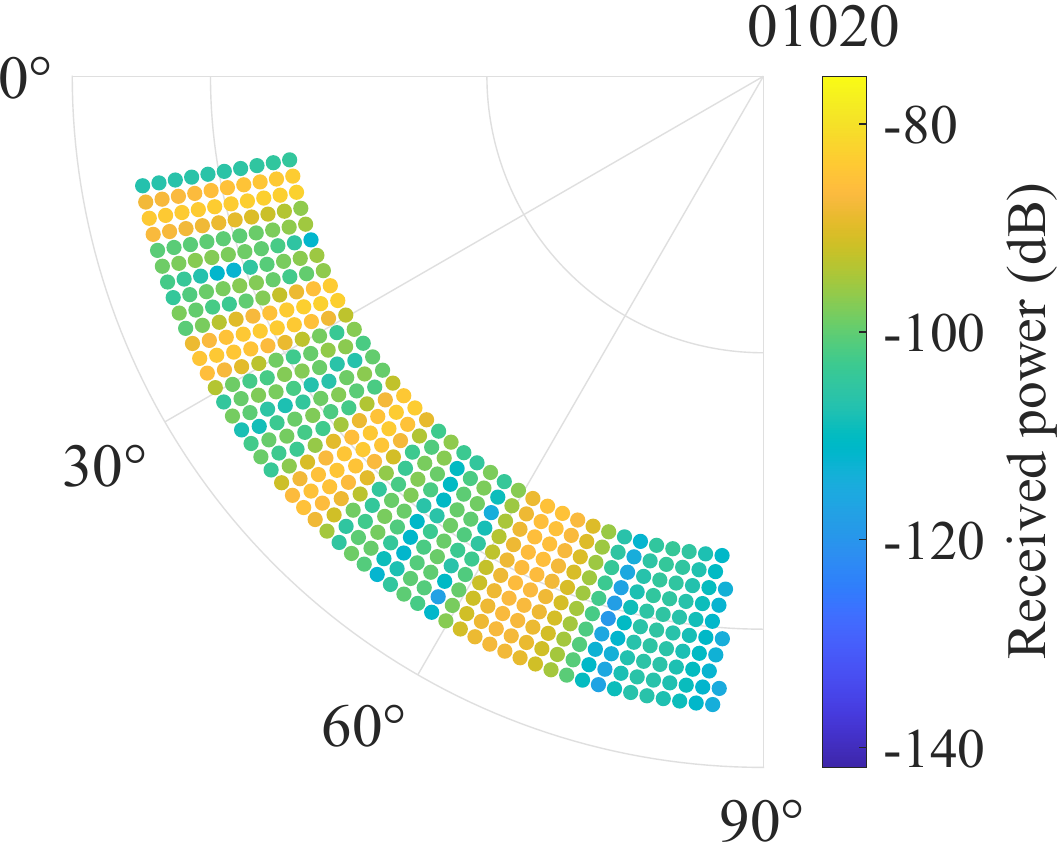}}
\subfigure[\label{Fig:4d}]{\includegraphics[width=0.241\textwidth]{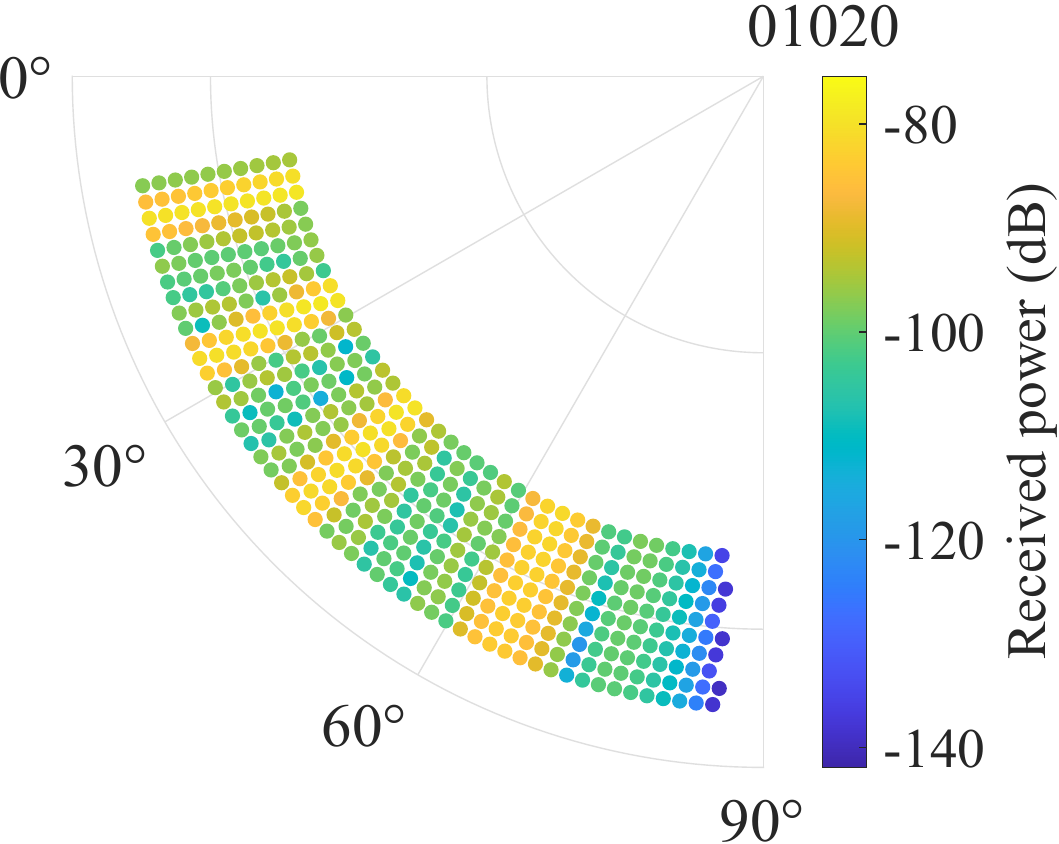}}
\subfigure[\label{Fig:4e}]{\includegraphics[width=0.241\textwidth]{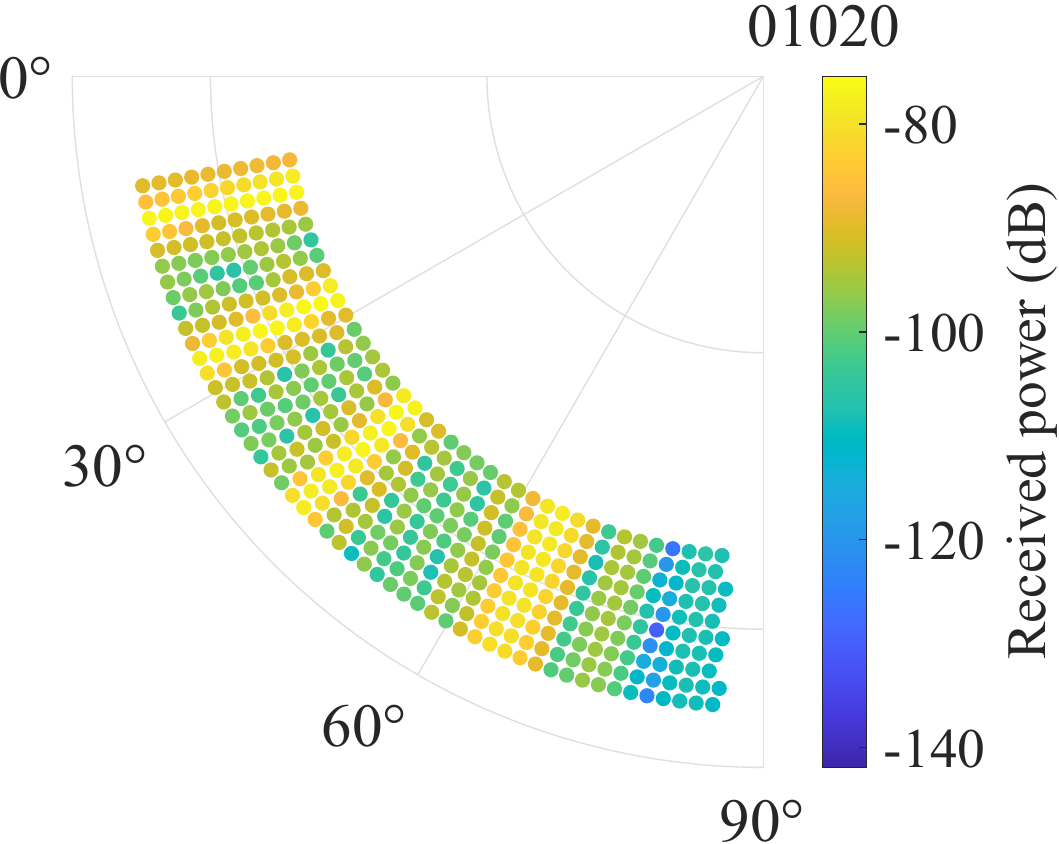}}
\caption{Received power vs. the angles and distances between the RXs and RIS for different RIS sizes without reflections. (a) RIS size $32 \times 32$, (b) RIS size $48 \times 48$, (c) RIS size $64 \times 64$, (d) RIS size $80 \times 80$, (e) RIS size $96 \times 96$. \label{Fig:rxpower_dist_ang_los}}
\vspace{-1.5em}
\end{figure}

To investigate whether the RX antennas at different locations benefit from the RIS, we simulate the LOS path from the TX to the RIS and from the RIS to each RX antenna for all four modes. Then we take the maximum received power for each RX antenna among all four modes. In this way, the RX antennas located at $13^\degree, 27^\degree, 43^\degree,$ and $65^\degree$ should all receive strong power due to the RIS assistance. The received power for all the users at different angles and distances from the RIS is plotted in Fig.~\ref{Fig:rxpower_dist_ang_los}. Figures~\ref{Fig:4a}, \ref{Fig:4b}, \ref{Fig:4c}, \ref{Fig:4d}, and \ref{Fig:4e} display the results for the $32\times32$, $48\times48$, $64\times64$, $80\times80$, and $96\times96$ RIS sizes, respectively. The radius of the polar plot is the distance between the RX and the RIS. The RIS is located at point $0$ in these figures. The color of these figures' curves represents the received power. 

From Fig.~\ref{Fig:rxpower_dist_ang_los} we can observe that the RXs located at $13^\degree, 27^\degree, 43^\degree,$ and $65^\degree$ receive the highest power. The RXs at other angles receive lower power because there is no strong reflection from the RIS in those angle ranges. With an increased distance between the RIS and the RX antenna, the received power is slightly reduced. However, since the distance change is not so much from $17.4$~m to $22.8$~m, the power reduction is not so significant. The maximum received power at the RXs increases from $-92.4$~dB to $-74.5$~dB with the RIS size increasing from $32\times32$ to $96\times96$. 

\begin{figure}[t]
\centering	
\subfigure[\label{Fig:5a}]{\includegraphics[width=0.241\textwidth]{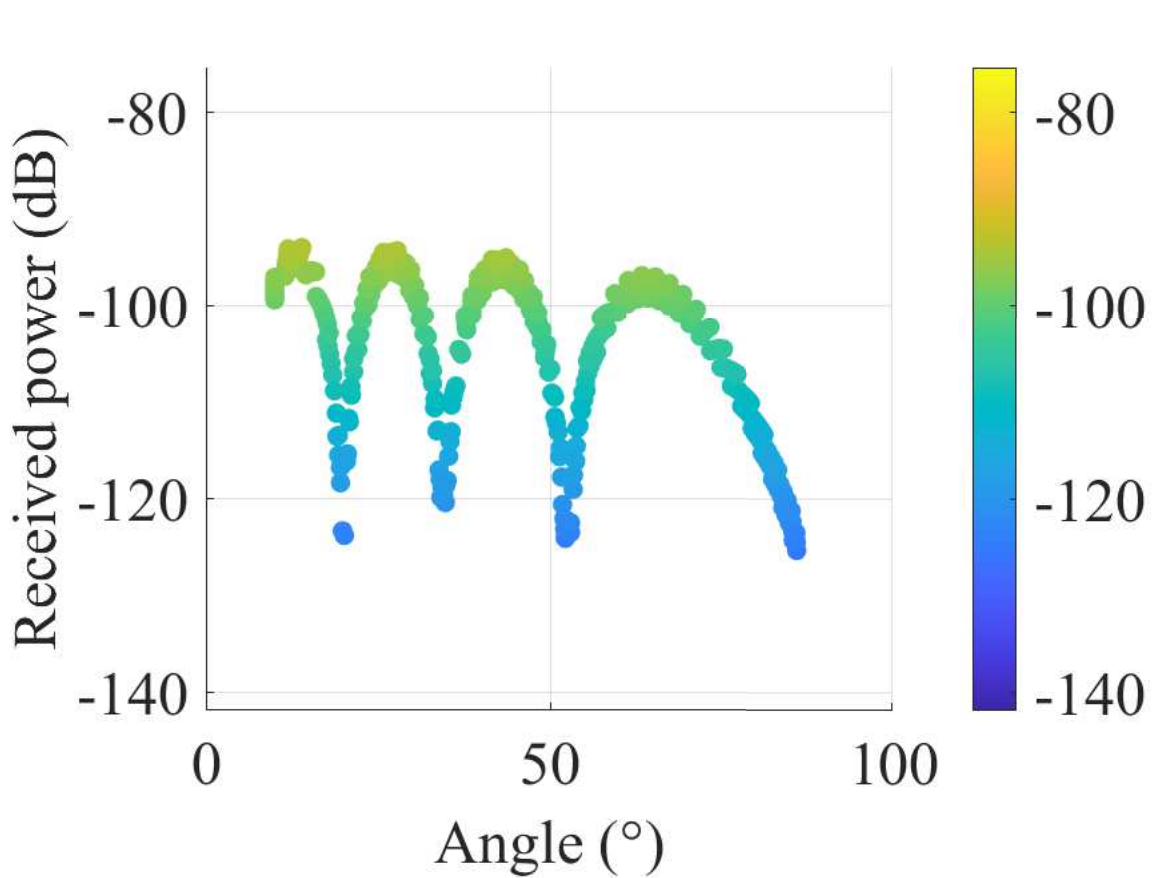}}
\subfigure[\label{Fig:5b}]{\includegraphics[width=0.241\textwidth]{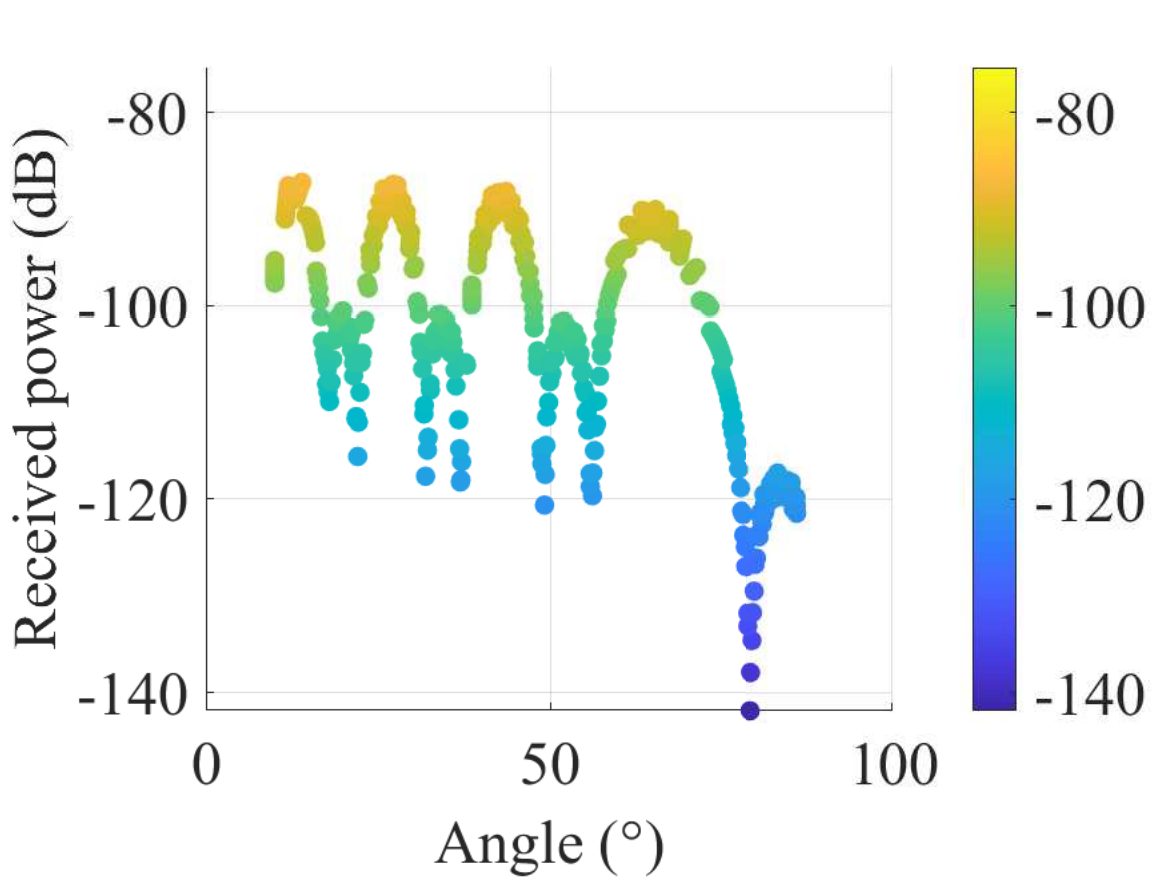}}
\subfigure[\label{Fig:5c}]{\includegraphics[width=0.241\textwidth]{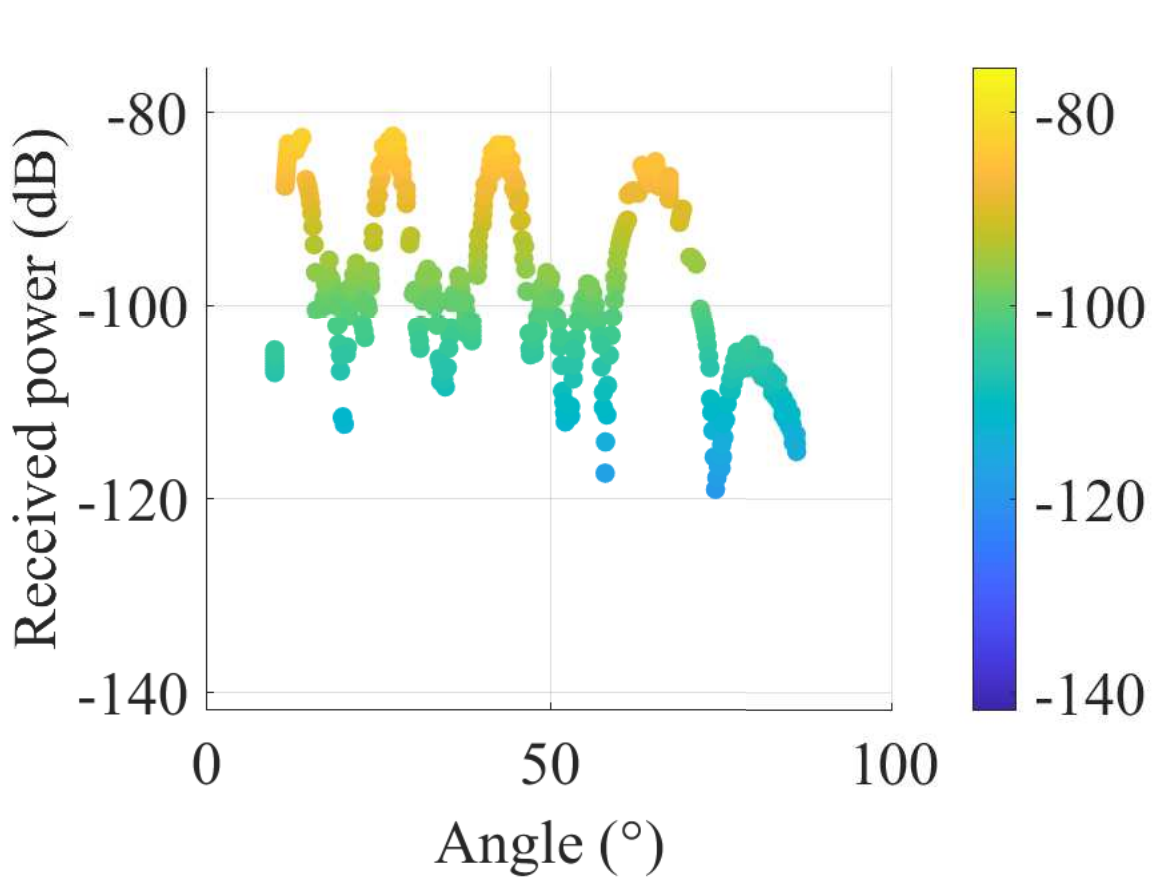}}
\subfigure[\label{Fig:5d}]{\includegraphics[width=0.241\textwidth]{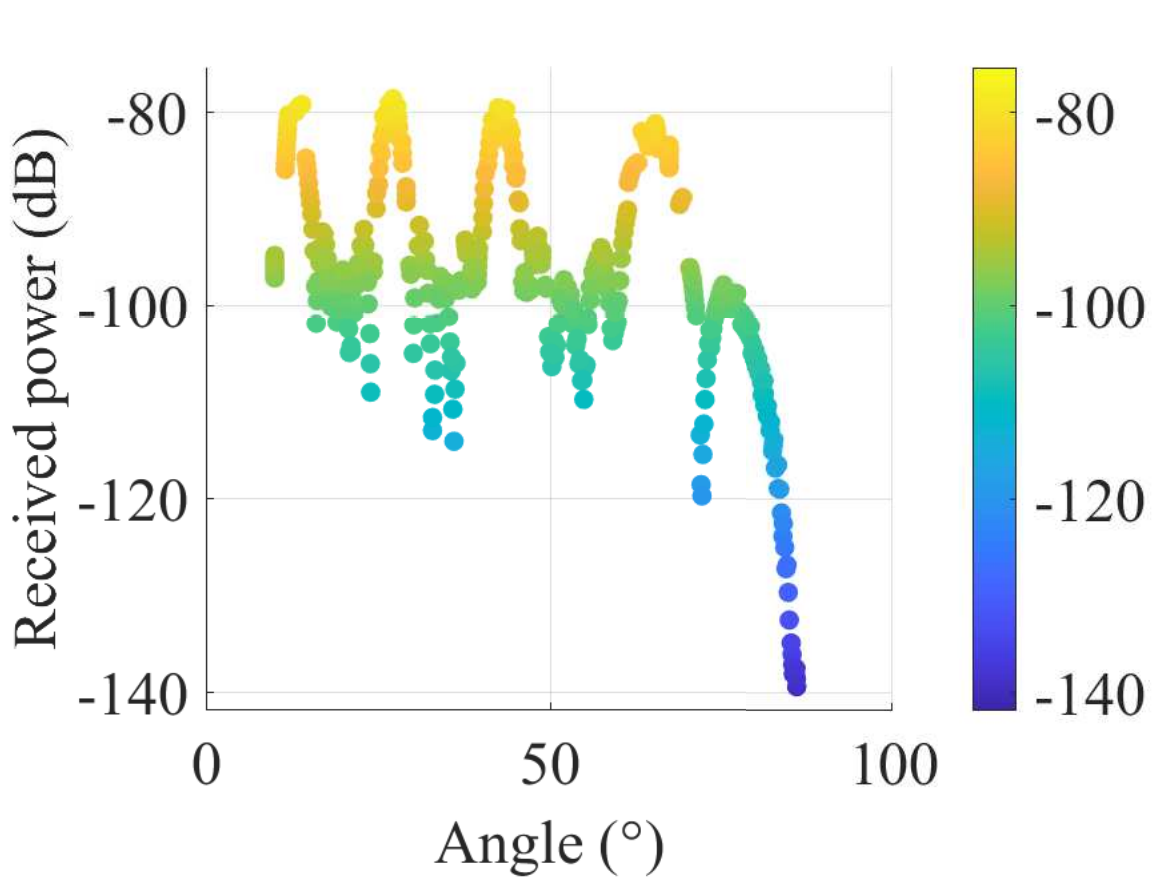}}
\subfigure[\label{Fig:5e}]{\includegraphics[width=0.241\textwidth]{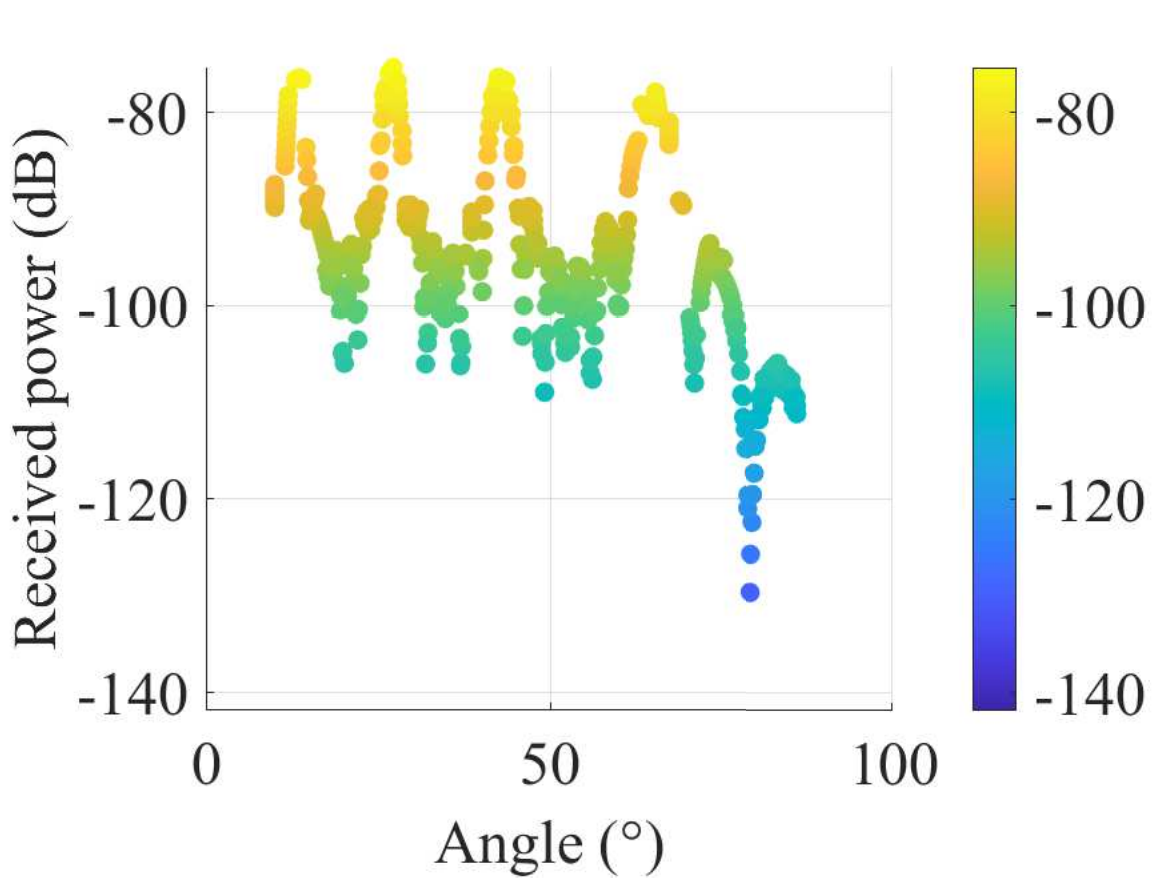}}
\caption{Received power vs. the angles between the users and the RIS for different RIS sizes without reflections. The scattering pattern of the RIS is seen in these figures.  (a) RIS size $32 \times 32$, (b) RIS size $48 \times 48$, (c) RIS size $64 \times 64$, (d) RIS size $80 \times 80$, (e) RIS size $96 \times 96$. \label{Fig:rxpower_ang_los}}
\end{figure}

To have a more detailed look at the received power at different angles, we plot the received power versus the angle results in Fig.~\ref{Fig:rxpower_ang_los}. At each angle, there are multiple points representing multiple RX antennas at that angle with different distances from the RIS. From Fig.~\ref{Fig:5a} to \ref{Fig:5e} are the results for  $32\times32$ to $96\times96$ RIS sizes, respectively. It is obvious that the received power at the RXs forms four strong beams at the four RIS reflection directions, which is related to the RIS scattering pattern. When the RIS size increases, the scattering pattern of the RIS at each reflection angle also becomes more directive and stronger. 

\begin{figure}[t]
\centering		
\subfigure[\label{Fig:9a}]{\includegraphics[width=0.241\textwidth]{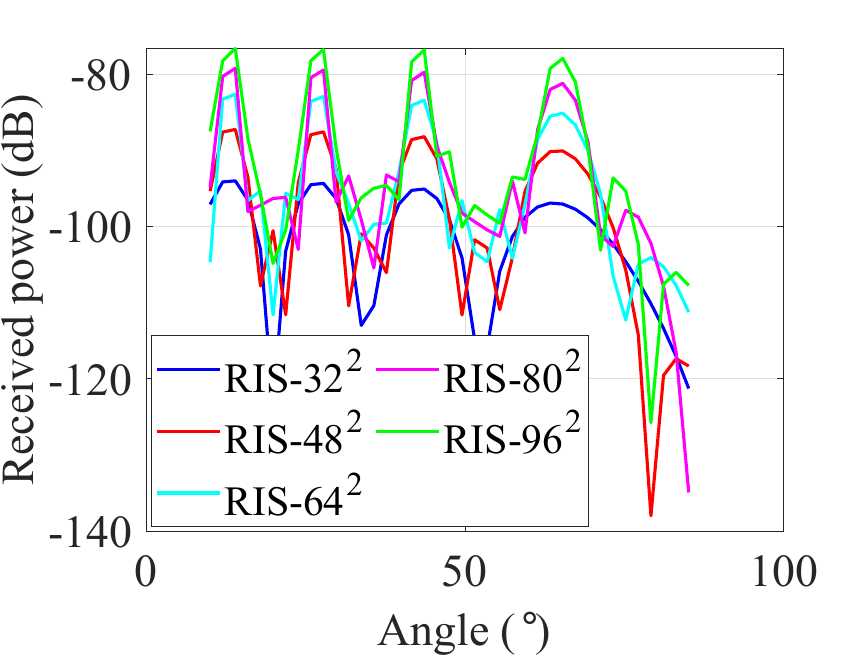}}
\subfigure[\label{Fig:9b}]{\includegraphics[width=0.241\textwidth]{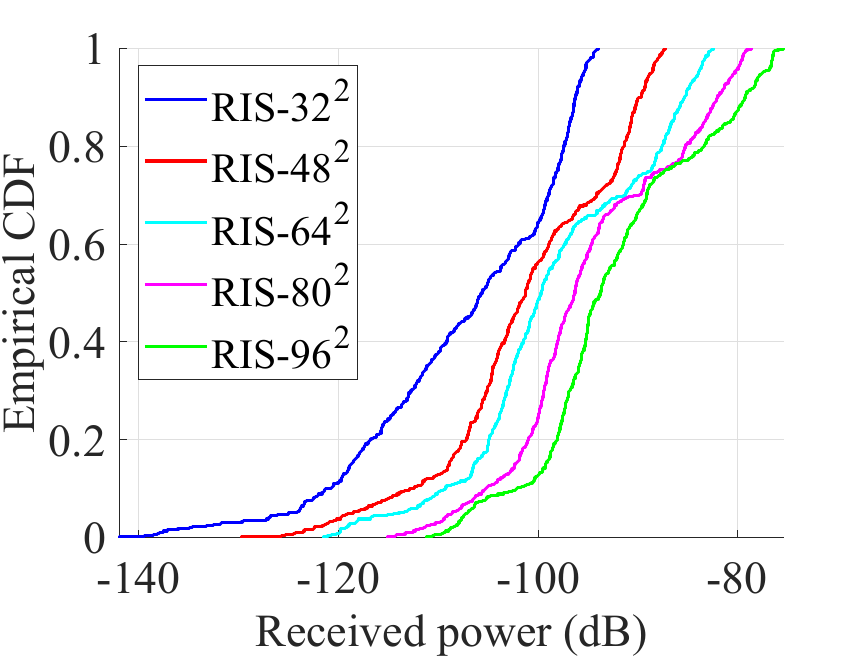}}
\caption{Received power comparison for different RIS sizes without reflections. (a) Received power vs. the angles toward the users at the same distances to the RIS, (b) ECDF of the received power comparison.\label{Fig:ecdfplot_los}}
\end{figure}

Next, we choose  $39$ RX antennas that are on the first arc at the $17.4$~m distance from the RIS. The received powers at the RX antennas for five RIS sizes are plotted in Fig.~\ref{Fig:9a}. The calculated differences of received powers at different RIS reflection angles are consistent with the results for the SISO scenario in Sec.~\ref{subsec:RIS_SISO_Scenario}. When the angle toward the RX antenna is not at one of the RIS reflection angles, the received power of these RX antennas is much lower. The differences between the five RIS sizes at those angles are also not very significant. The \ac{ECDF} results are plotted in Fig.~\ref{Fig:9b} to compare the overall received power at all the RX antennas for the five RIS sizes. 

\subsection{Multi-user Scenario with Multi-path Propagation}
\label{sec:MPRTsimulation}
In this section, we consider a multi-path propagation scenario. The setup is the same as in Sec.~\ref{sec:LOSRTsimulation}, except that in this scenario we include reflection paths. 
It should be noted that since our RIS is designed for illumination at normal incidence and reflections into a set of four angles, it can be effectively used only for these paths and for the reciprocal ones. The RIS scattering patterns for illuminations from other directions need to be calculated separately (for RIS realized as periodical arrays, this issue is considered in \cite{Rubio2021}). For simplicity, here we consider only one LOS path for the TX-RIS link, but three reflection paths for the RIS-RX link. To investigate the difference of the received power at the RIS between the LOS path and reflection paths, we run simulations with $0$, $3$, and $6$ reflections, and 
find that the difference between the LOS and $3$ or $6$-reflections paths is smaller than $1$~dBW, which is very small. Hence, even though it is not so realistic to assume only one LOS path between the TX and the RIS, it is still reasonable to use this assumption for simulations. 

\begin{figure}[t]
\centering		
\subfigure[\label{Fig:6f}]{\includegraphics[width=0.241\textwidth]{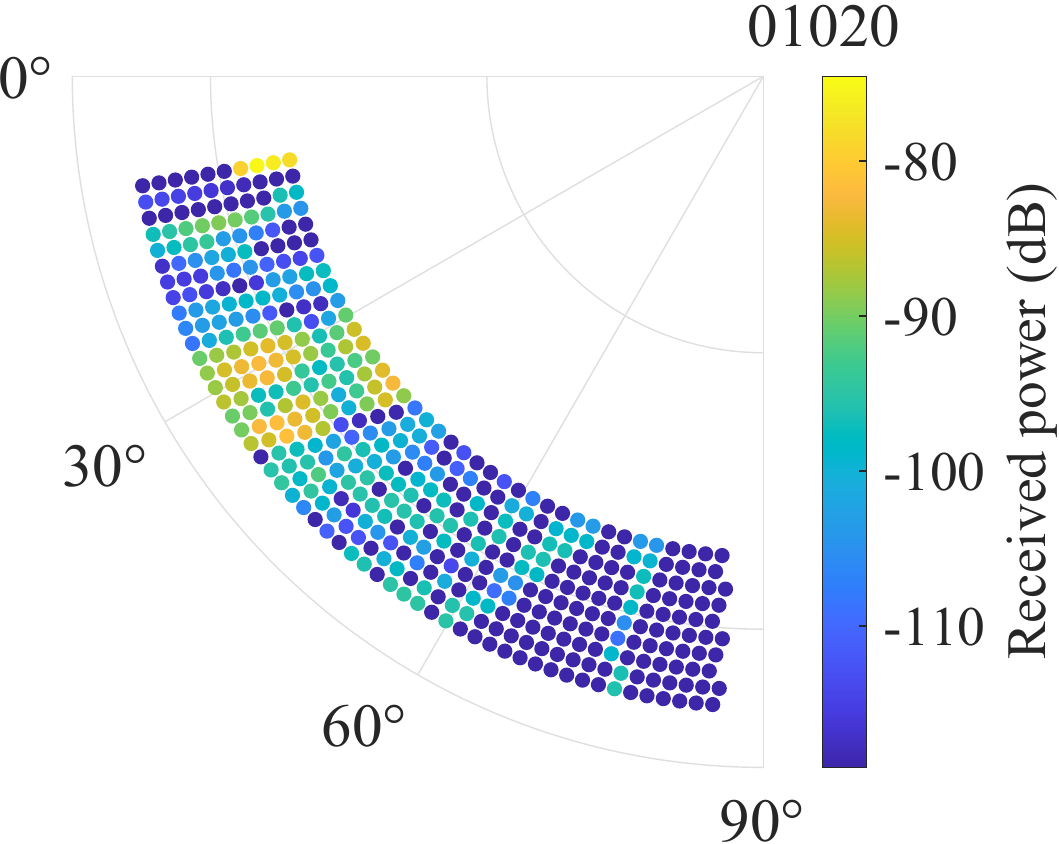}}
\subfigure[\label{Fig:6a}]{\includegraphics[width=0.241\textwidth]{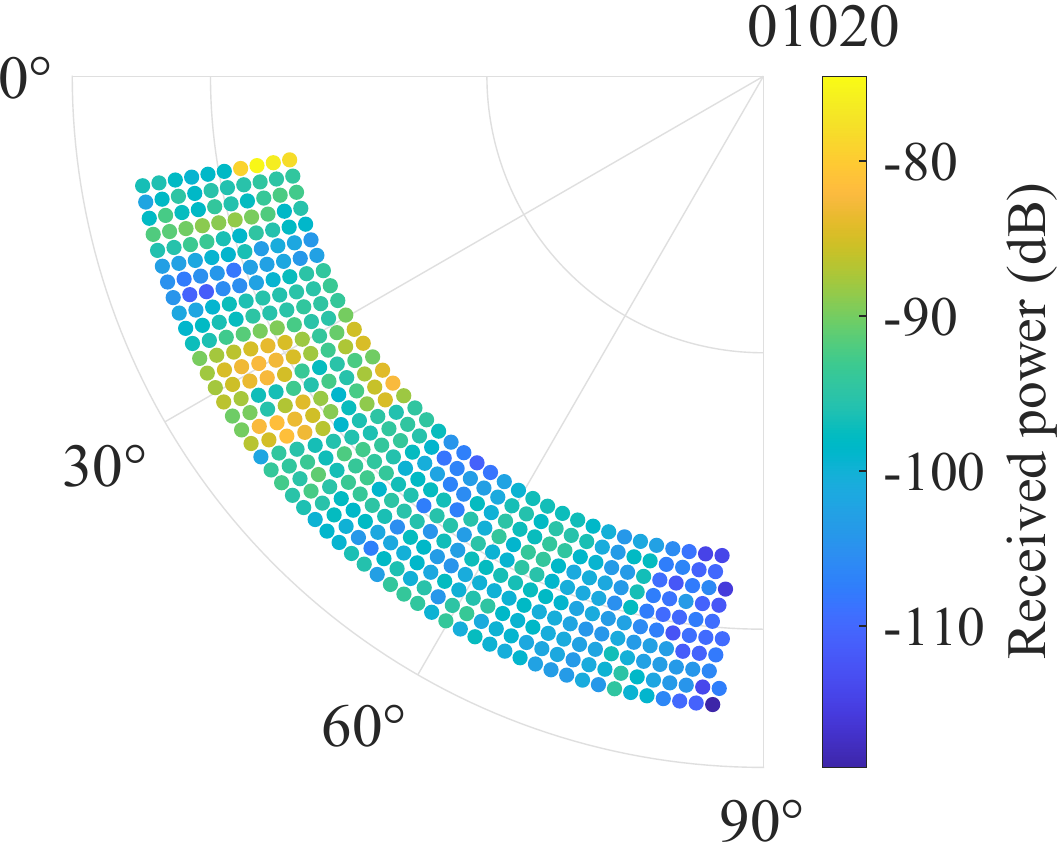}}
\subfigure[\label{Fig:6b}]{\includegraphics[width=0.241\textwidth]{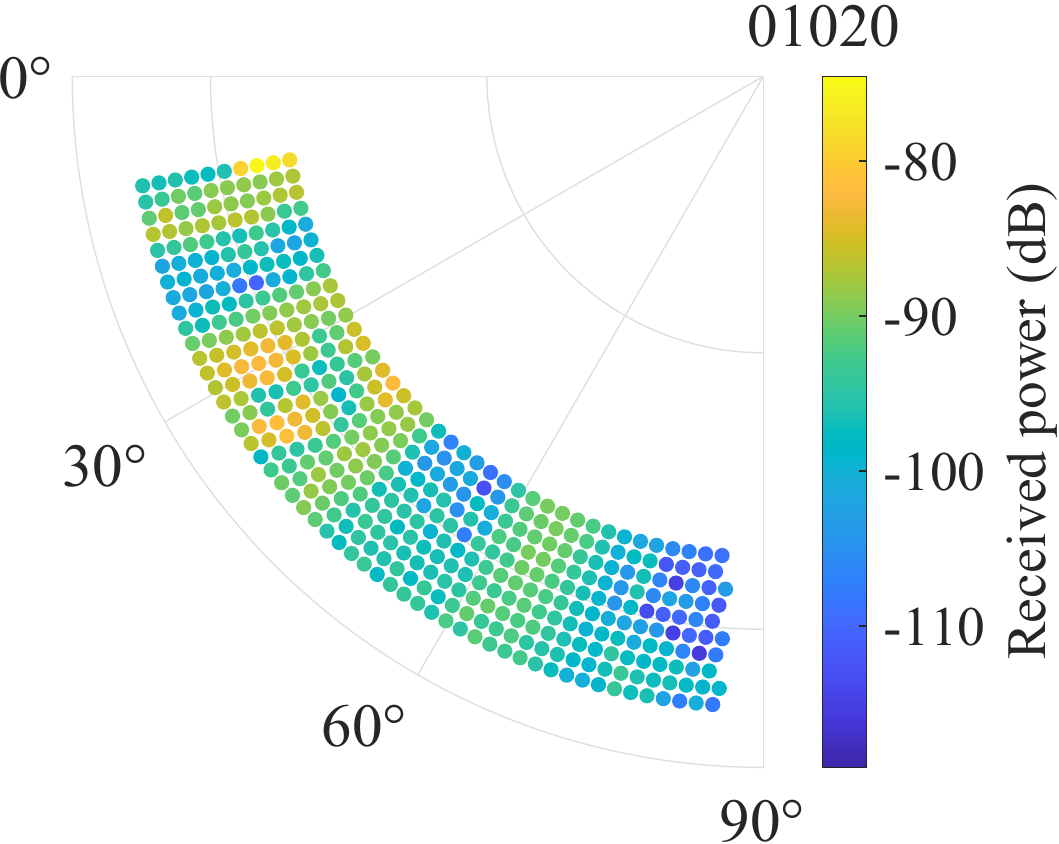}}
\subfigure[\label{Fig:6c}]{\includegraphics[width=0.241\textwidth]{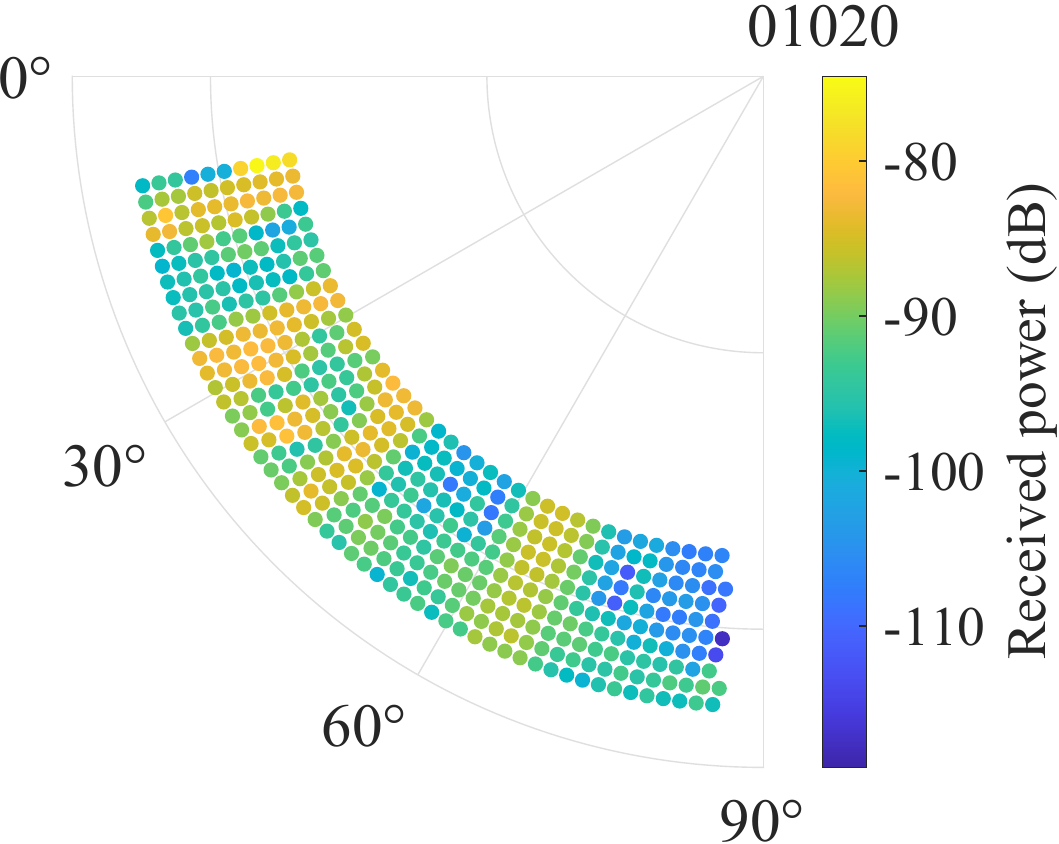}}
\subfigure[\label{Fig:6d}]{\includegraphics[width=0.241\textwidth]{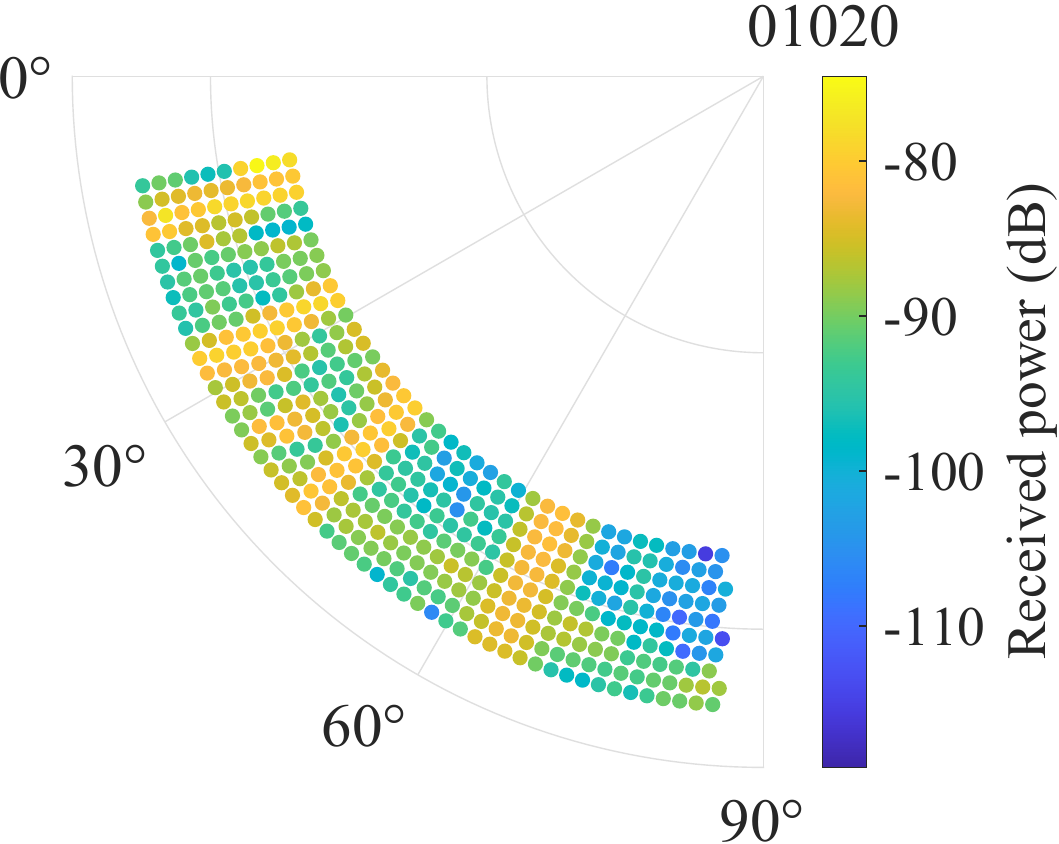}}
\subfigure[\label{Fig:6e}]{\includegraphics[width=0.241\textwidth]{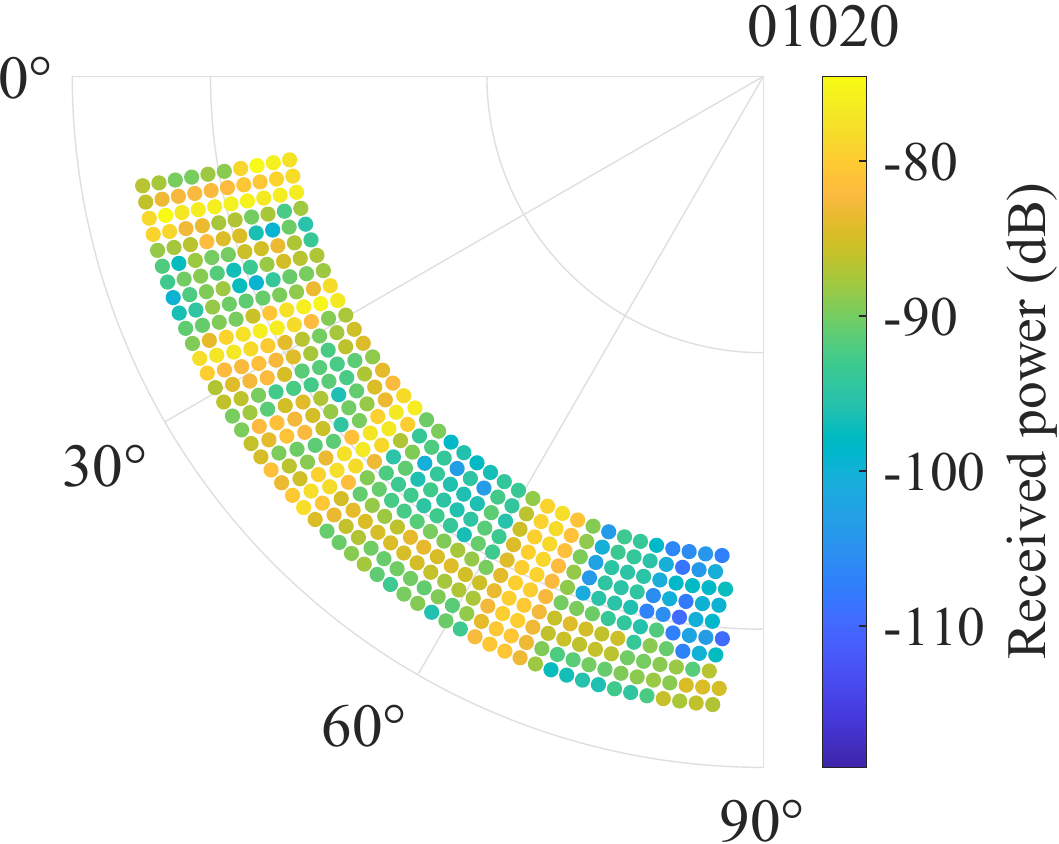}}
\caption{Received power vs. angles and distances between the users and the RIS of different sizes with $3$ reflections. (a) Without RIS, (b) RIS size $32 \times 32$, (c) RIS size $48 \times 48$, (d) RIS size $64 \times 64$, (e) RIS size $80 \times 80$, (f) RIS size $96 \times 96$. \label{Fig:rxpower_dist_ang_mp}}
\end{figure}

The received power results versus the distances to the RX antenna and the angles are shown in Fig.~\ref{Fig:rxpower_dist_ang_mp}. In addition to the results with five RIS sizes that are shown in Figs.~\ref{Fig:6a} -- \ref{Fig:6e}, the results without RIS are plotted in Fig.~\ref{Fig:6f}. From these figures, we notice that when there is no RIS, only some RX antennas located at the angle of $10^\degree$ and in the range of $[26^\degree ~ 41^\degree]$ receive relatively high power, while many RX antennas receive only noise. However, when including a RIS in this scenario, almost all RX antennas are covered and receive a significant amount of signal power.

\begin{figure}[t]
\centering		
\subfigure[\label{Fig:7f}]{\includegraphics[width=0.241\textwidth]{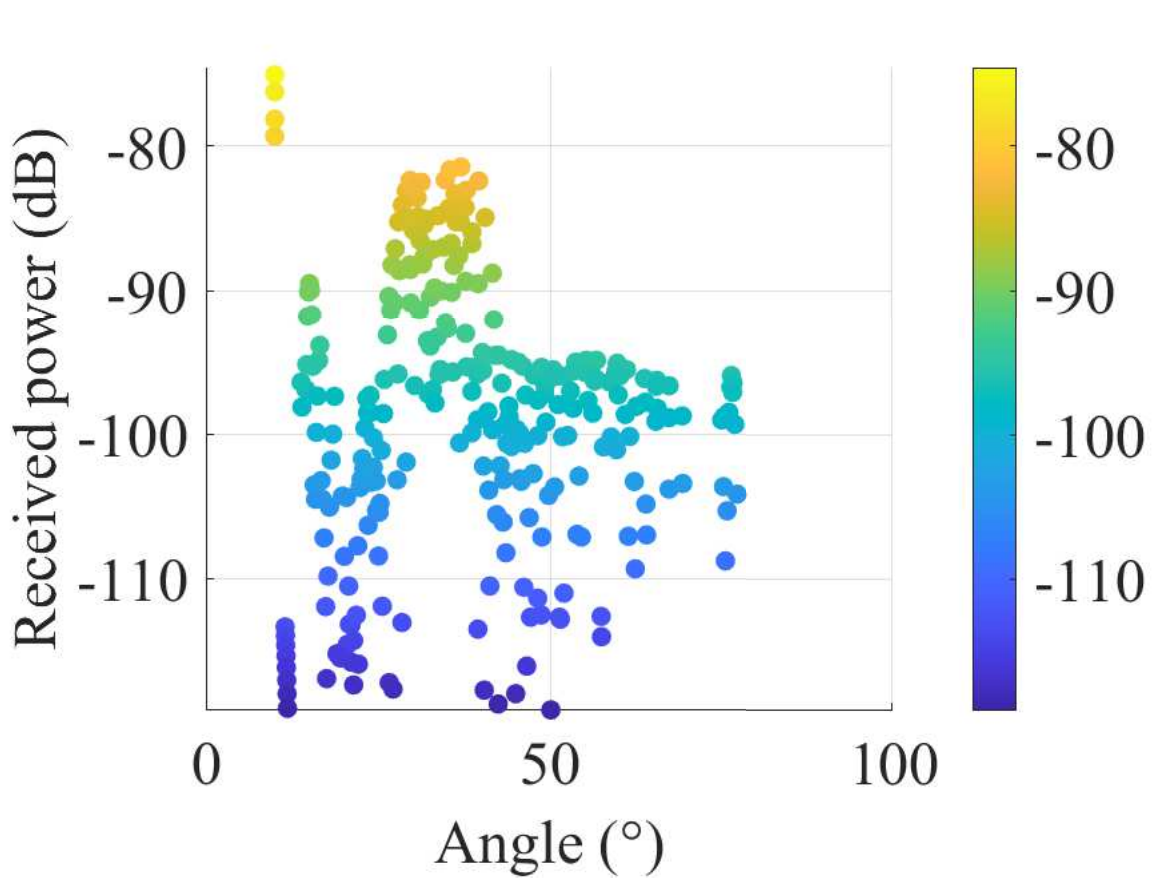}}
\subfigure[\label{Fig:7a}]{\includegraphics[width=0.241\textwidth]{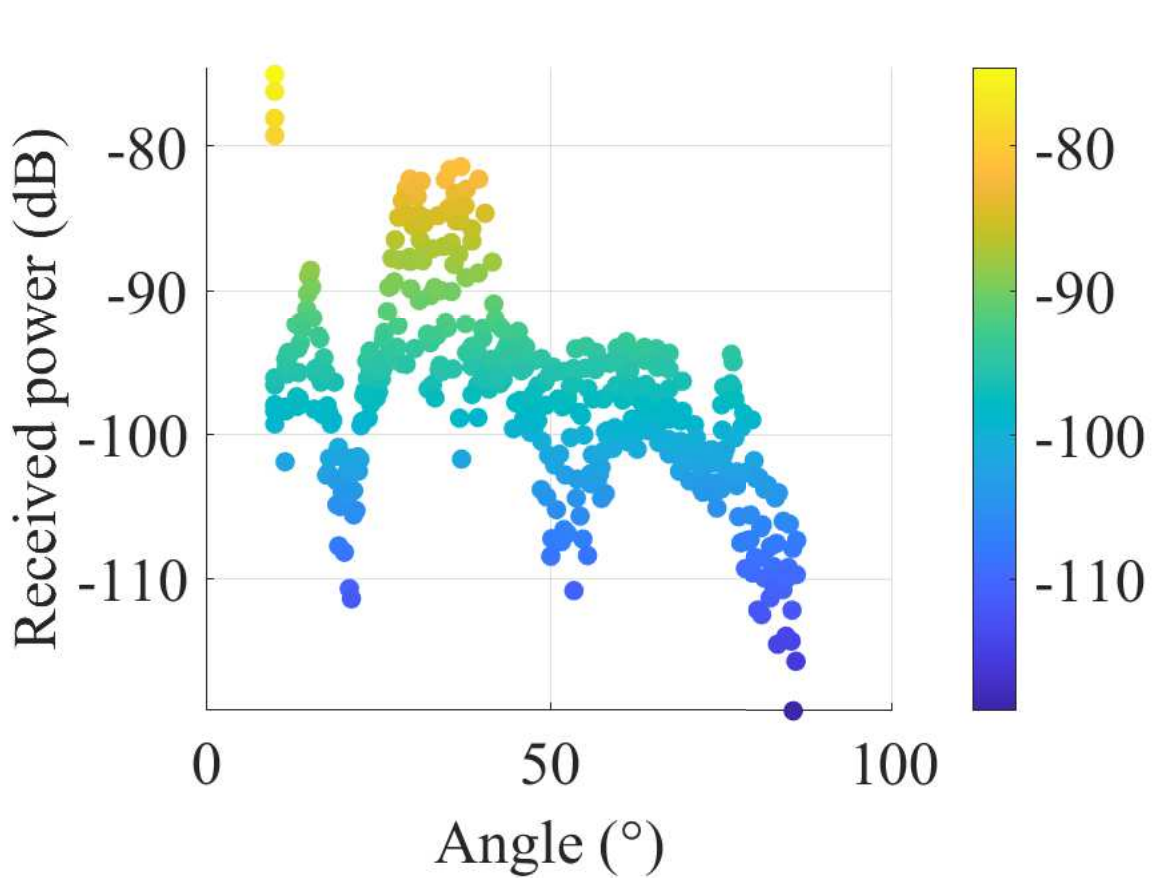}}
\subfigure[\label{Fig:7b}]{\includegraphics[width=0.241\textwidth]{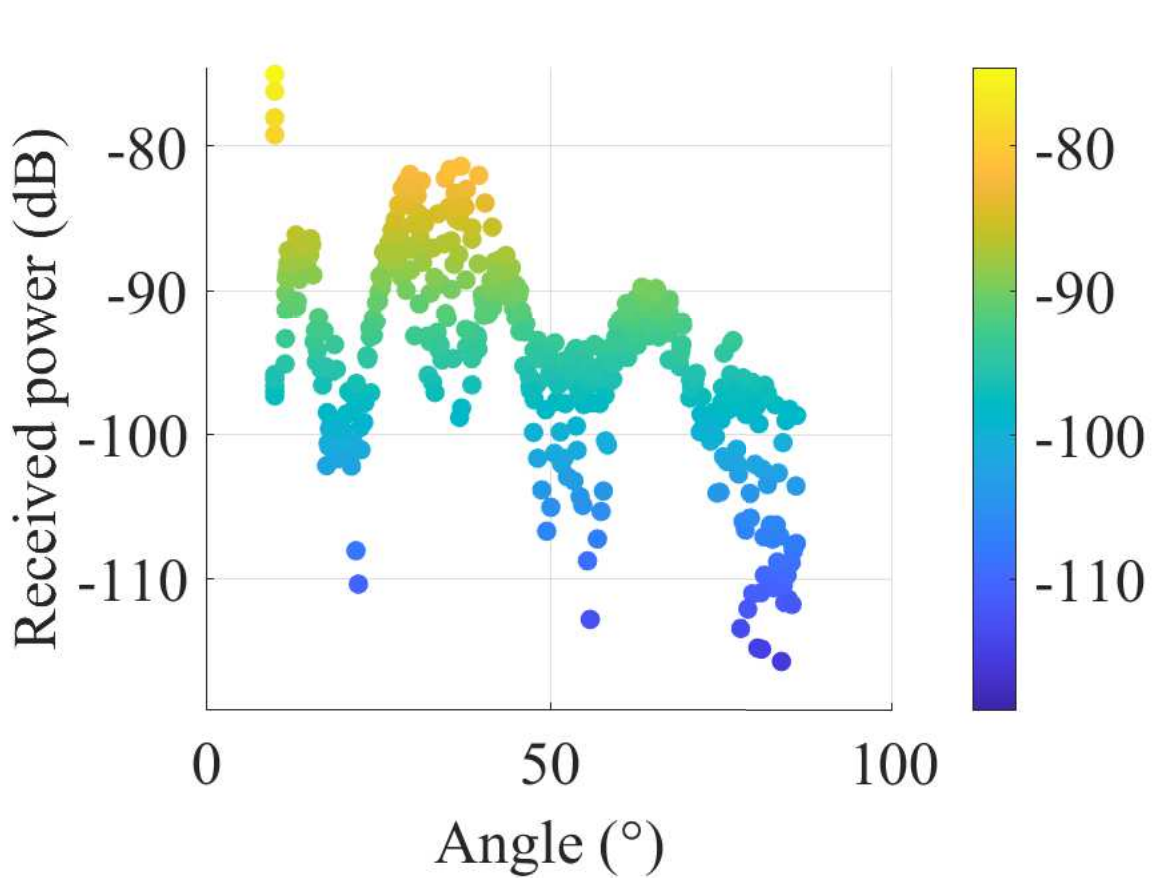}}
\subfigure[\label{Fig:7c}]{\includegraphics[width=0.241\textwidth]{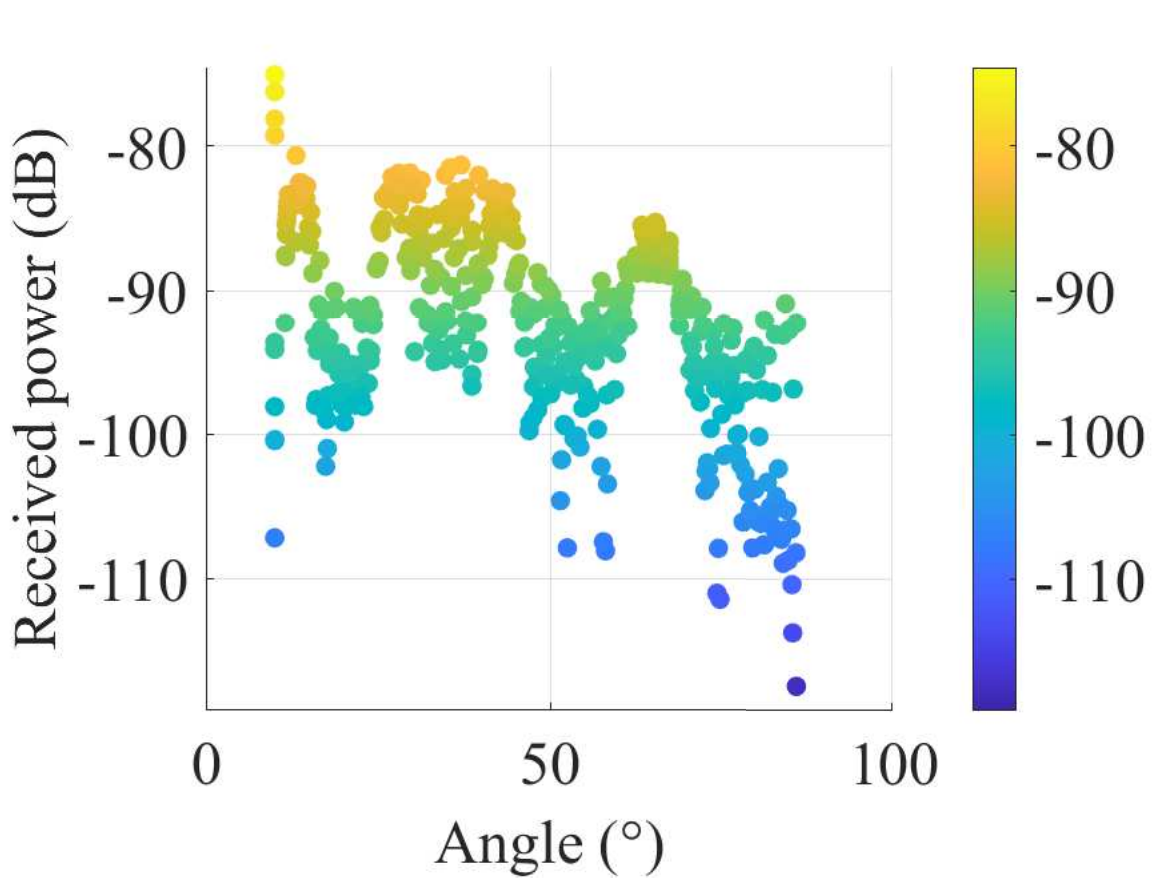}}
\subfigure[\label{Fig:7d}]{\includegraphics[width=0.241\textwidth]{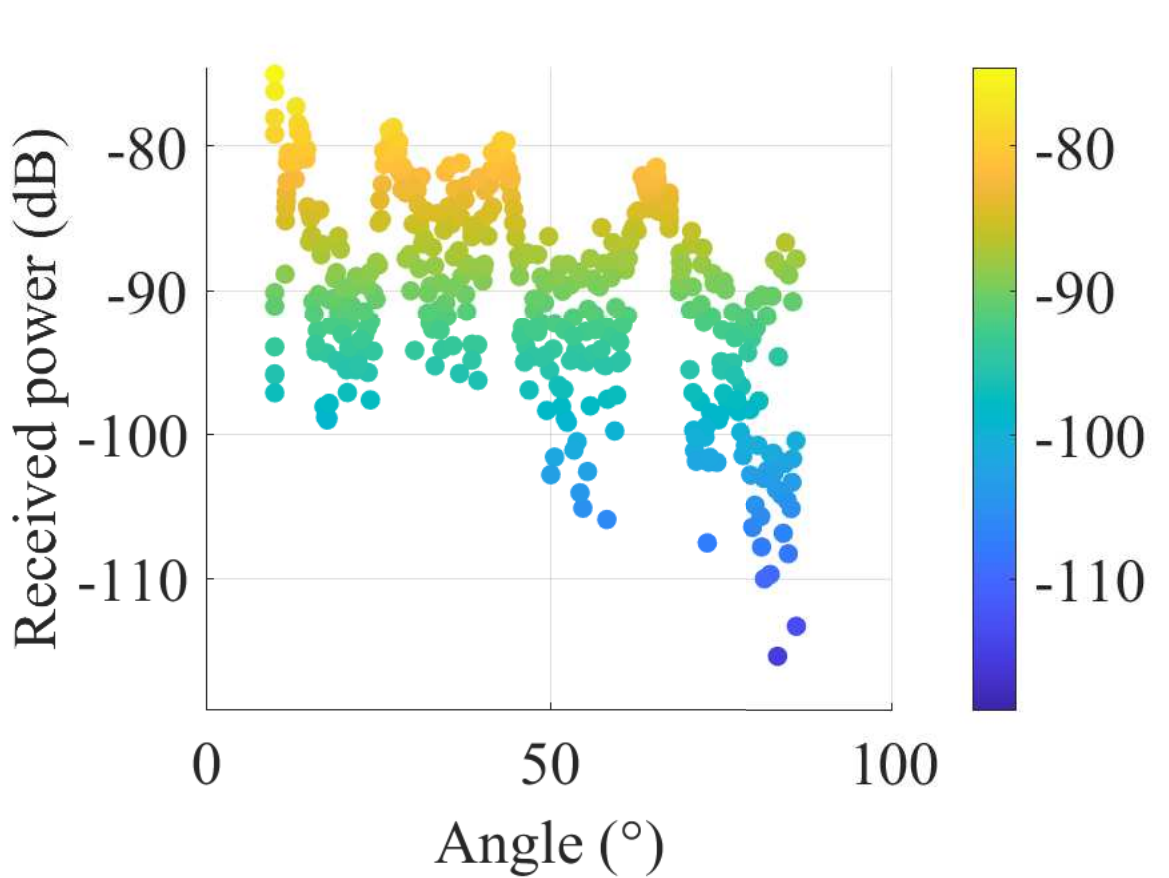}}
\subfigure[\label{Fig:7e}]{\includegraphics[width=0.241\textwidth]{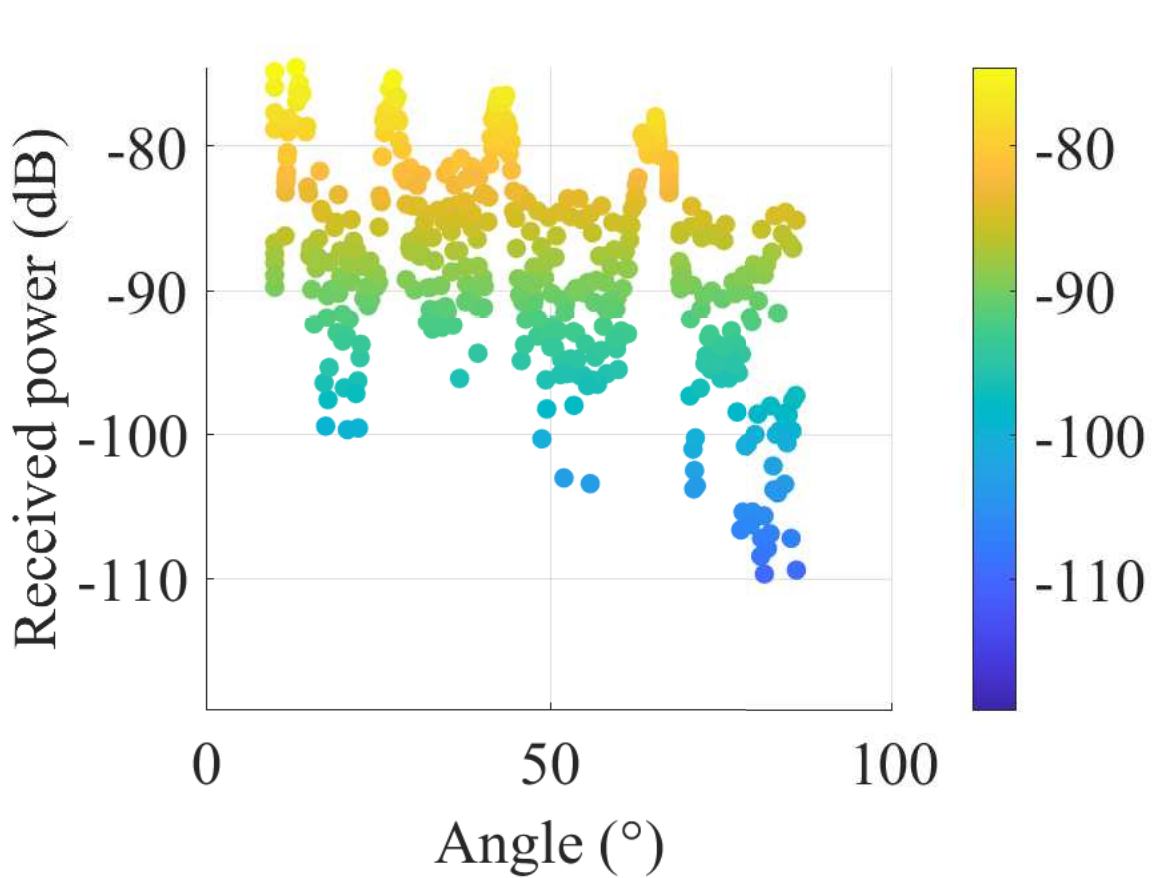}}
\caption{Received power vs. angles of the users and the RIS of different sizes with $3$ reflections. (a) Without RIS, (b) RIS size $32 \times 32$, (c) RIS size $48 \times 48$, (d) RIS size $64 \times 64$, (e) RIS size $80 \times 80$, (f) RIS size $96 \times 96$. \label{Fig:rxpower_ang_mp}}
\end{figure}

The received power as a function of the angle is plotted in Fig.~\ref{Fig:rxpower_ang_mp}. Compared to the LOS scenario in Fig.~\ref{Fig:rxpower_ang_los}, the received powers at the users that are not located along the RIS reflection directions have significant improvement in the multi-path scenario. For example, in the LOS scenario with the RIS size $96\times96$, the received powers at the RX antennas in the $[50^\degree ~ 57^\degree]$ range are in the range of [$-109$~$-96$]~dB. However, in the multi-path scenario, the received powers at those RX antennas are in the [$-104$~$-83$]~dB range and are higher than in the LOS scenario on average. It is obvious that after multiple reflections from the walls, floor, ceiling, and cabinet, the transmitted signals have a good chance of reaching those RX antennas in blind spots.

\begin{figure}[t]
\centering		
\subfigure[\label{Fig:8a}]{\includegraphics[width=0.241\textwidth]{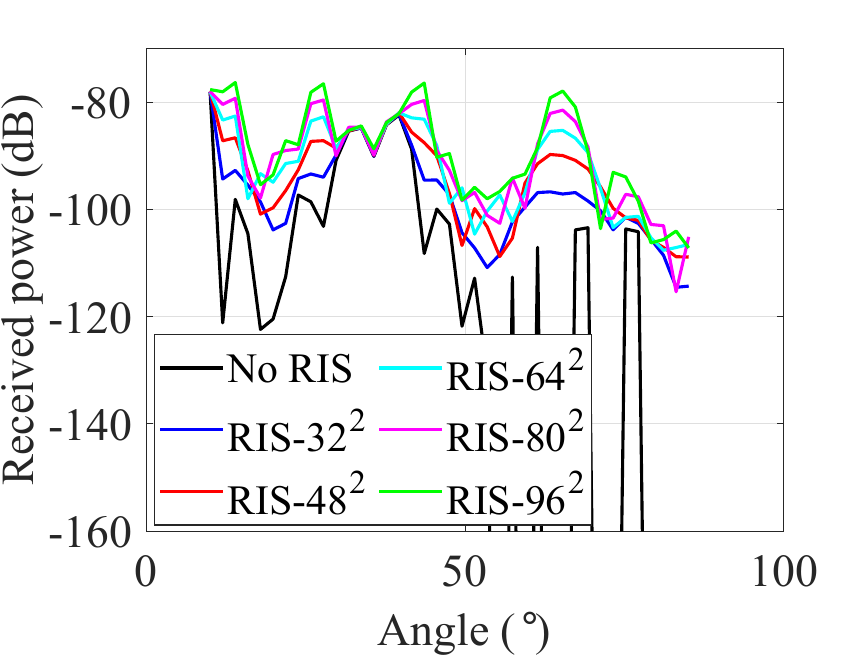}}
\subfigure[\label{Fig:8b}]{\includegraphics[width=0.241\textwidth]{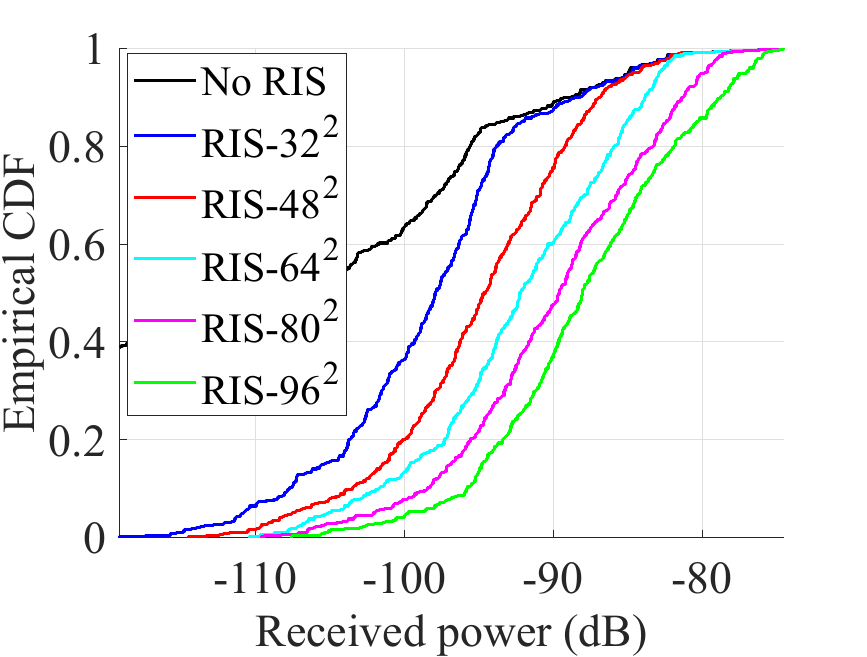}}
\caption{Received power comparison with different RIS sizes with $3$ reflections. (a) Received power vs. angles of the users which have the same distances to the RIS, (b) ECDF of the received power comparison.\label{Fig:ecdfplot_mp}}
\end{figure}

Similarly, in Fig.~\ref{Fig:8a} we compare the results for RIS of five sizes with $39$ RX antennas at the first arc that are $17.4$~m away from the RIS. Compared to the case without RIS, the scenarios with RIS lead to significant improvement of received power at these RXs, especially at the RIS reflection directions and with a larger RIS size. The ECDF results of the received power are plotted in Fig.~\ref{Fig:8b}. The curve for the case of 'No RIS' does not start from zero because many RXs receive just noise, and they are not included in this figure.
\begin{figure*}[t]
\centering	
\includegraphics[width=0.9\textwidth]{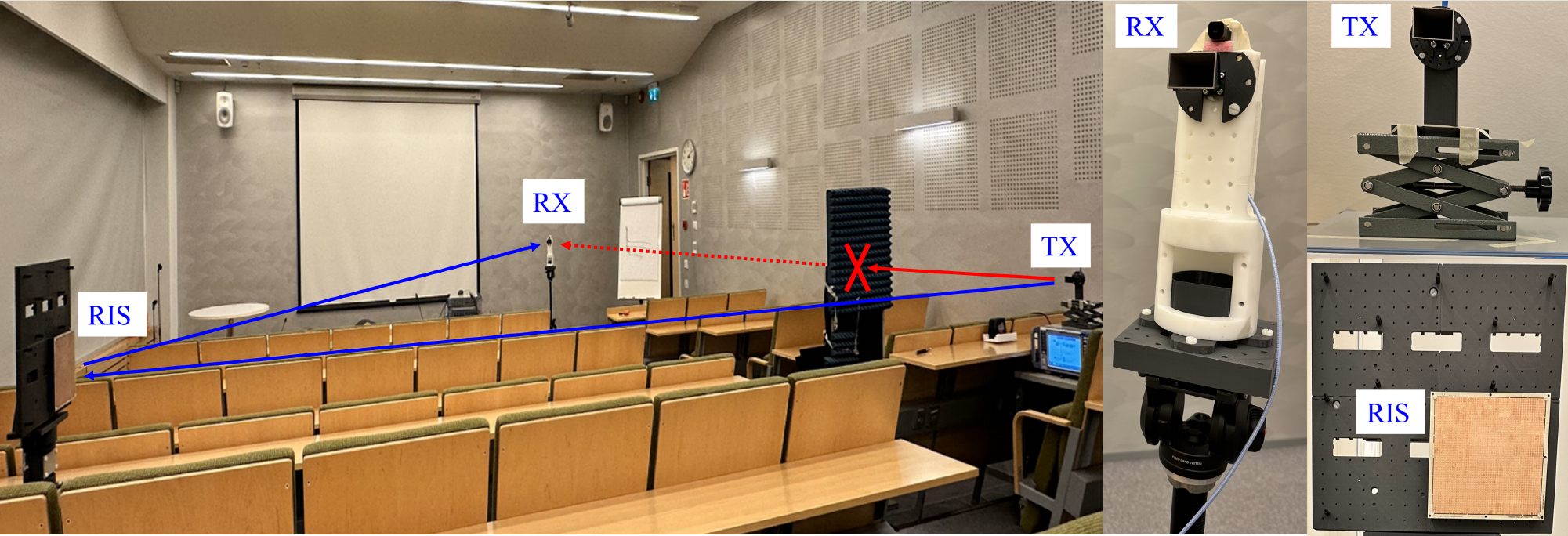}
\caption{Measurement setup in the auditorium with the $48\times48$-sized static AR prototype, TX, and RX antennas.}\label{Fig:measure_audi}
\vspace{-1em}
\end{figure*}

\begin{figure*}[t]
\centering		
\subfigure[\label{Fig:room3D}]{\includegraphics[width=0.52\textwidth]{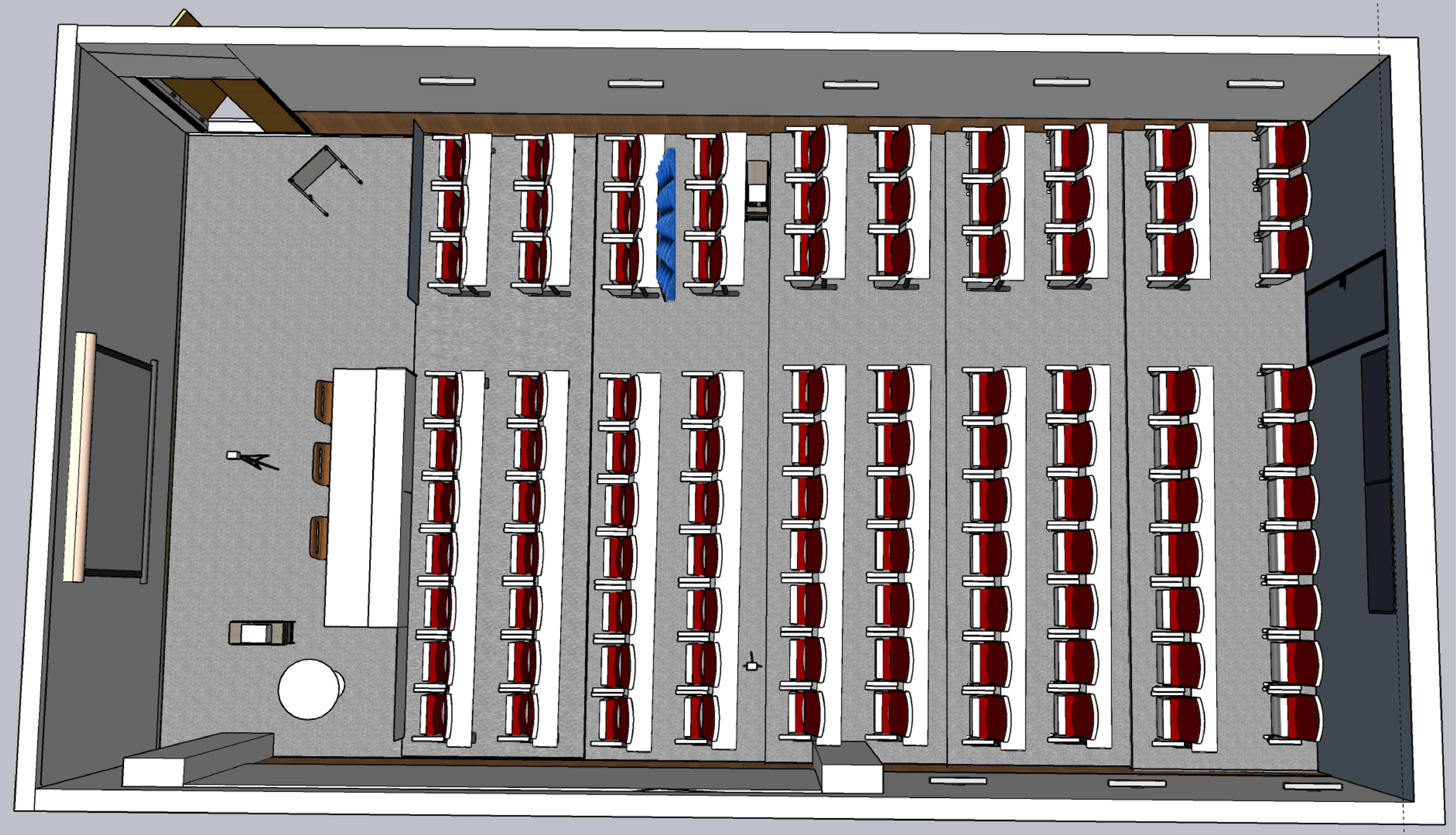}}
\subfigure[\label{Fig:roomRT}]{\includegraphics[width=0.38\textwidth]{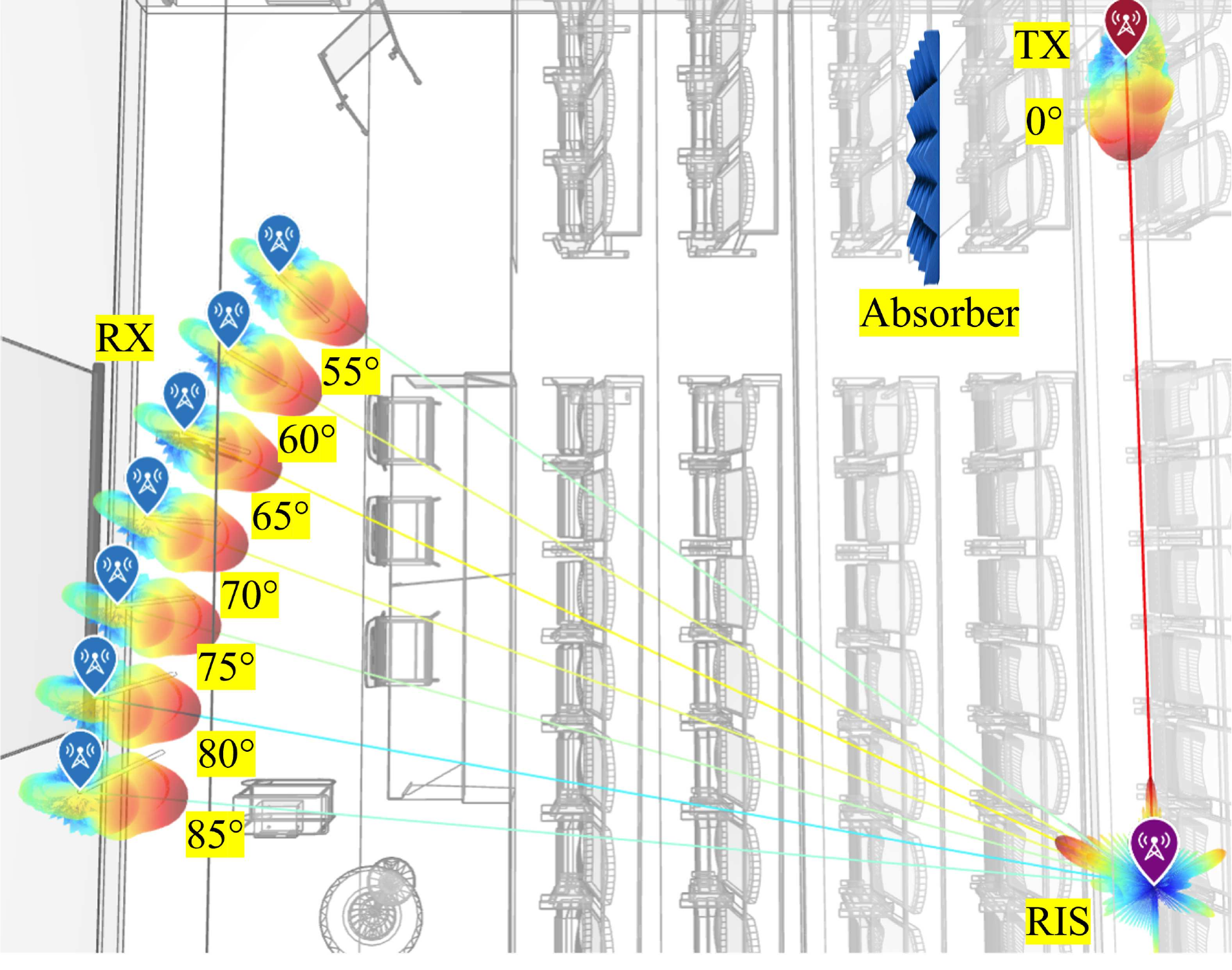}}
\caption{Ray tracing model of the auditorium. (a) Top view of the 3D model of the auditorium, (b) Ray tracing in the auditorium model.\label{Fig:RT_audi}}
\end{figure*}

In summary, the strategy of implementing a RIS as an antenna in a ray tracer is proved to be correct, which can be also applied to other ray tracers. The ray tracing simulation results from a SISO and multi-user LOS scenario prove that the maximum received powers at the RX antennas fulfill the power scaling law~\cite{Wu2018}, which is actually from the communication theory where it just considers the RIS element number and phase shifts and does not involve any EM properties of the RIS, but our RIS model is from EM perspective and modeled as a whole antenna. So far we have reached a good agreement when applying the communication theory to a realistic RIS from the EM perspective. In addition to the comparison between the scenario with multiple reflections and without reflections, we can conclude that the contributions of a RIS are highly dependent on its reflection directions, the RIS sizes, and the reflections from the environment. The user located at the RIS reflection angles can receive the maximum power, the bigger RIS size also contributes more power to the user, and reflection paths in the environment can contribute to the user coverage improvement.

\section{Experimental Results}\label{sec:RISExperimental}
In Part I we introduced a prototype of a static $48\times48$ array and measured the scattering pattern of it in an anechoic chamber. In this part we performed over-the-air measurements at $26$~GHz with the same prototype in an auditorium at Nokia Bell Labs Espoo office, to test the communication link performance and our ray tracing model with the realistic \ac{AR}. 

\subsection{Indoor Measurement and Ray Tracing Settings}
\label{subsec:RISmeas_audi}
The measurement scenario is shown in Fig.~\ref{Fig:measure_audi}. The dimensions of the auditorium are $14\times8\times3$~(m$^3$). The TX and RX antennas are the same horn antennas with a maximum gain of $G_t = G_r = 18$~dBi and a beam width of $22\degree$ at $26$~GHz. The height of the TX, RX, and the AR is $1.5$~m, the distance between the TX and AR is $R_1=5.5$~m, and the distance between the AR and the RX is $R_2=7$~m. The TX horn antenna is connected to a vector signal generator via a cable, the RX horn antenna is connected to a low noise amplifier (LNA), and then connected to a signal \& spectrum analyzer via cables. The signal generator is connected to the signal \& spectrum analyzer through a reference clock and Ethernet cable for synchronizing the signals. The TX cable loss is $L_t = 2.5$~dB, the sum of the LNA gain and the RX cable loss is $G_a = 19.9$~dB. We use a 400-MHz channel bandwidth and $16$~QAM modulated 5G NR wavemode for transmitted signals, and the transmitted power at the TX side is $P_t = 6$~dBm. 

To measure the TX-AR-RX link, we use a wave absorber to block the direct link between the TX and the RX antennas. The TX and the AR are fixed and are facing each other. The RX antenna is placed at $55^\circ$, $60^\circ$, ..., $85^\circ$ of the AR, respectively, but the distance between each RX location and the AR is always $7$~m. We orient the direction of the RX antenna to always face the AR at each location. In the end, we obtain $7$ different received power values at the RX antenna from the $7$ different locations. We denote this power as $P_m$ in dBm.

To simulate the same measurement scenario in a ray tracer, we first create a 3D model using SketchUp and import it to a ray tracer. The top view of the model is displayed in Fig.~\ref{Fig:room3D}. This 3D model replicates the real dimensions of the whole room and the objects inside it. For practical reasons, the MATLAB ray tracer suits the auditorium scenario better than alternative ray tracers. Therefore, we use the MATLAB ray tracer in this section for simulations. 

We model a horn antenna in MATLAB and use it for TX and RX antennas in ray tracing simulations. The maximum gain of the horn antenna is $18$~dBi. Then we use a similar way of implementing RIS as in the Wireless InSite to implement the realistic \ac{AR} in the MATLAB ray tracer. The locations of the TX, RX antennas, and the \ac{AR} are the same as in the measurement. Figure~\ref{Fig:roomRT} shows the ray tracing of the TX-AR LOS link and the AR-RXs LOS links. We first set the reflection number as $0$ to observe the received power from the AR-assisted LOS links, and compare it with the theoretical results obtained from Sec.~\ref{sec:SLSsimulation}, since the two methods in Sec.~\ref{sec:SLSsimulation} also considers only the LOS paths. Then we set $3$ reflections in ray tracing simulations for the TX-AR and AR-RX links to include the reflection paths from the room objects and compare the results with measurement results, since the reflections cannot be ignored in a realistic environment. We denote the simulated powers at $7$ locations as $P_\text{r,orig}$. Then, considering the cable losses and the LNA gain, we obtain $P_\text{RT,orig} = P_\text{r,orig} - L_t + G_a $.

\subsection{Results Comparison between the Theoretical Model, Ray Tracing, and Measurement}\label{sec:resultscompare}

\begin{figure}[t]
\centering		
\subfigure[\label{Fig:origresults}]{\includegraphics[width=0.241\textwidth]{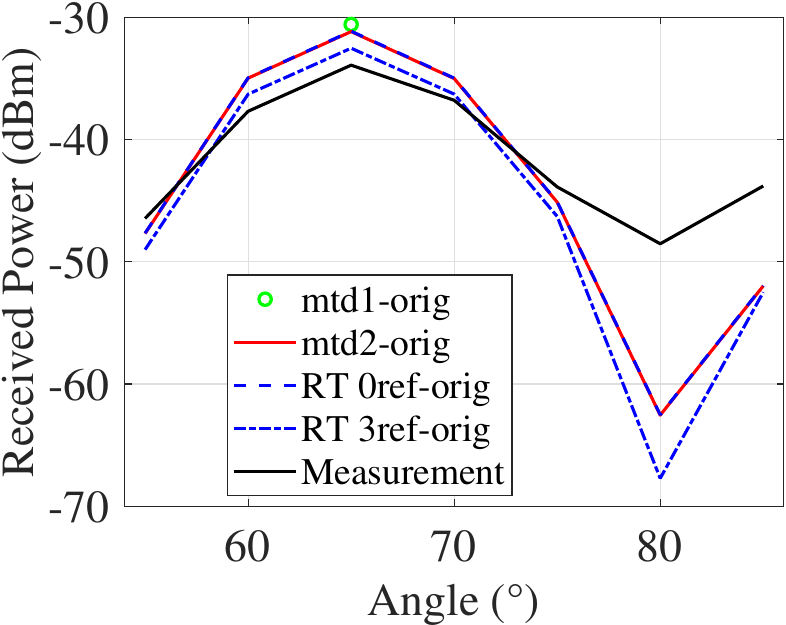}}
\subfigure[\label{Fig:correctresults}]{\includegraphics[width=0.241\textwidth]{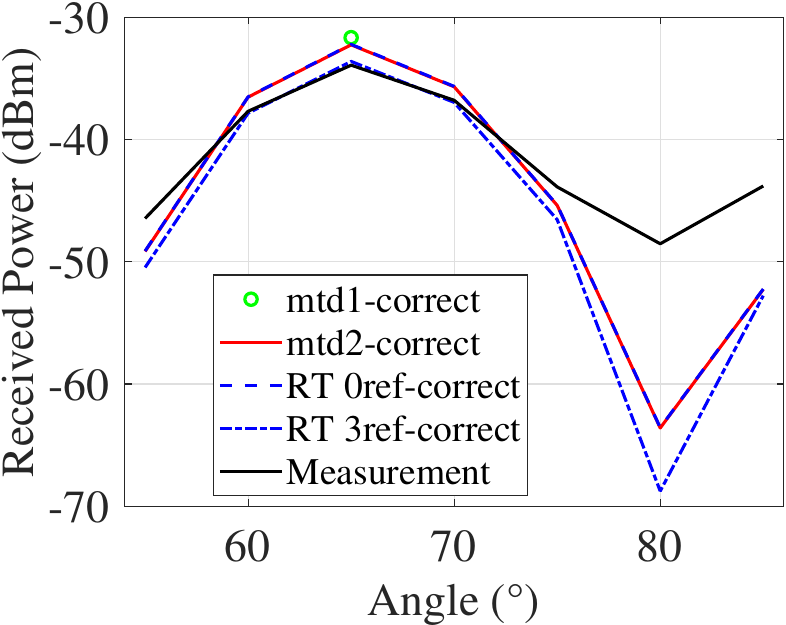}}
\caption{Received power vs. angle results comparison between theory, ray tracing simulation and measurement. (a) Original results from method $1, 2$, ray tracing simulations and measurements, (b) Corrected results from method $1, 2$, ray tracing simulations, and measurement results.\label{Fig:pwRIS48_26g}}
\end{figure}

In this section, we first compare the received power results between the two methods from Sec.~\ref{sec:SLSsimulation}, the measurement result $P_m$, and the ray tracing simulation result $P_\text{RT,orig}$ as shown in Fig.~\ref{Fig:origresults}. It is worth noting that method $1$ can only give the results at the RIS targeted direction, it cannot be used to calculate received powers at other directions that the RIS is not designed for. In our case, we only consider the received power at $65^\circ$ with method $1$ since this AR is designed for $65^\circ$. We denote this result as $P_{\text{mtd1,orig}}$. With method $2$ and ray tracing, we obtain received powers also at other angles 
by utilizing the respective RIS radiation patterns from CST simulations. The result from method $2$ is denoted as $P_{\text{mtd2,orig}}$.
From Fig.~\ref{Fig:origresults} we observe that the ray tracing results with zero reflection are the same as from method $2$, and they are very close to the theoretical result from method $1$, which is consistent with our analysis in Sec.~\ref{sec:2mtdcompare} and Sec.~\ref{subsec:RIS_SISO_Scenario}. However, the ray tracing results with $3$ reflections have about $1.4$~dB difference compared to measurement results at $60^\circ$ and $65^\circ$.

To investigate whether this difference is from our ray tracing model or from the measurement system, we perform a reference measurement for the TX-RX link and compare it with theoretical results. In this reference measurement, we do not include the AR and the absorber, but let the TX and RX antennas directly face each other every time when we move the RX antenna to the $7$ locations. The results from this measurement is denoted as $P_{m,\text{LOS}}$ in dBm. Because it is very easy to validate this kind of LOS scenario through theory, i.e., we calculate the free space path loss between the two antennas using the equation $P_\text{FS} = P_t G_t G_r \left(\frac{\lambda}{4\pi R}\right)^2$ in W, where $R$ is the distance between the TX and RX antennas. Then adding the cable losses and LNA gain, we obtain a theoretical received power $P_\text{theory} = P_\text{FS} - L_t + G_a$ in dBW, where the $P_\text{FS}$ here is in dBW. We find the power differences between the theoretical value and the measurement results are very small: $P_\text{diff} = P_\text{theory} - P_{m,\text{LOS}} = [1.5, 1.6, 1.1, 0.7, 0.3, 1.0, 0.3]$~dBm for the angles of $[55^\circ,60^\circ,65^\circ,70^\circ,75^\circ,80^\circ,85^\circ]$, respectively. This difference may be due to the system loss in our measurement setups, including alignment errors of the antennas, and is not included in the theoretical model.

Now if we take into account the $P_\text{diff}$ when comparing the simulations and measurement results, i.e., use this $P_\text{diff}$ to correct the theoretical and simulation results and obtain $P_{\text{mtd1,correct}} = P_{\text{mtd1,orig}} - P_\text{diff}$, $P_{\text{mtd2,correct}} = P_{\text{mtd2,orig}} - P_\text{diff}$, and $P_\text{RT,correct} = P_\text{RT,orig} - P_\text{diff}$. The comparison with measurement results $P_m$ is shown in Fig.~\ref{Fig:correctresults}. 
We find the measurement results at $60^\circ$, $65^\circ$, and $70^\circ$ now agree very well with the ray tracing results with $3$ reflections. We can thus conclude that our designed RIS works as we expected. It is also proved that our theoretical analysis for a RIS-assisted link is correct, the 3D auditorium model is accurate, the RIS implementation in Wireless InSite and in the MATLAB ray tracer is correct, the theoretical analysis and ray tracing methods work not only for a perfect RIS, but also for a realistic lossy \ac{AR}.

\section{Conclusion}\label{sec:conclusion}
In Part I we studied the scattering synthesis for multimode \ac{AR}s. Based on that, we designed a periodic \ac{AR} that supports multiple reflection directions. By optimizing the load impedances of each unit cell, with continuous and $1$ to $4$-bit resolution quantized reactive loads, the \ac{AR} can be configured to reflect waves toward one main direction while suppressing reflections in other directions. The designed lossless RIS can achieve almost perfect reflection efficiencies in the desired directions. The experiment with a manufactured finite reflector validated our design and reached a good agreement with CST simulation results. In this Part II, we used the designed perfect RIS and the manufactured reflector prototype to evaluate their \ac{EM} properties and communication performances. We implemented both the RIS and the AR into two different ray tracers and validated the results with the theory. In addition, we investigated the quantization effect on the RIS implementation and concluded that with a $3$-bit quantization resolution, the RIS can already achieve very good results. Furthermore, we analyzed the large-scale fading of RIS-assisted communication links through \ac{EM} simulation, system-level, and ray tracing simulations, as well as through indoor measurements in a room using a static \ac{AR} as a test vehicle. The results demonstrated that our RIS design from the \ac{EM} aspects, RIS implementation in the two ray tracers, and the system-level and ray tracing simulations with the two RISs from the communication aspects, are all correct and consistent with the measurement results. 

From the EM design perspective, we acknowledge that our designed multi-mode reflector is intended for reflection in discrete anomalous angles. The coverage between these angles is a topic of an extension to this work in the future. The unit cell loads of the multi-mode reflector would be made reconfigurable so that a non-perfect anomalous reflection in the gap angles would be allowed by the design, however improving the coverage between the discrete modal anomalous angles. Another possibility is to construct a multi-mode static anomalous reflector from multiple single-reflection angle sub-panels, i.e., not having a RIS as a reconfigurable surface, but multiple static anomalous reflectors side-by-side to implement the same functionality as the multi-mode RIS would do.

From the communication perspective, the validated connections between the EM and communication analysis for a RIS can accelerate the RIS technology realization. For example, by passing a limited set of macroscopic RIS parameters from the EM design to the RIS-tailored Vienna system-level simulator, the simulator can deal with realistic RISs in large-scale scenarios. In addition, by importing the radiation pattern of a RIS into ray tracers, one can simulate the RISs in different scenarios taking the propagation environment's effects into consideration. By using our approaches with the software, we can obtain a good estimation of the system performance with realistic RISs in realistic scenarios without the need to conduct measurements.

\section*{ACKNOWLEDGEMENT}

The authors express gratitude to Professor Do-Hoon Kwon from the University of Massachusetts Amherst, USA, for the valuable suggestions and discussions regarding array antenna scattering synthesis for periodic reflectors.

\ifCLASSOPTIONcaptionsoff
  \newpage
\fi
\bibliographystyle{IEEEtran}
\bibliography{IEEEabrv,Bibliography}

\end{document}

%% file: Acronyms.tex
\begin{acronym}[DSTTDSGRC]
\setlength{\itemsep}{-3pt}
\acro{BS}{base station}
\acro{ECDF}{empirical cumulative distribution function}
\acro{EM}{electromagnetic}
\acro{LOS}{line-of-sight}
\acro{SISO}{single-input single-output}
\acro{SLS}{system-level simulator}
\acro{AR}{anomalous reflector}
\end{acronym}

%% file: IEEE_TAP.bbl
\begin{thebibliography}{10}
\providecommand{\url}[1]{#1}
\csname url@samestyle\endcsname
\providecommand{\newblock}{\relax}
\providecommand{\bibinfo}[2]{#2}
\providecommand{\BIBentrySTDinterwordspacing}{\spaceskip=0pt\relax}
\providecommand{\BIBentryALTinterwordstretchfactor}{4}
\providecommand{\BIBentryALTinterwordspacing}{\spaceskip=\fontdimen2\font plus
\BIBentryALTinterwordstretchfactor\fontdimen3\font minus \fontdimen4\font\relax}
\providecommand{\BIBforeignlanguage}[2]{{%
\expandafter\ifx\csname l@#1\endcsname\relax
\typeout{** WARNING: IEEEtran.bst: No hyphenation pattern has been}%
\typeout{** loaded for the language `#1'. Using the pattern for}%
\typeout{** the default language instead.}%
\else
\language=\csname l@#1\endcsname
\fi
#2}}
\providecommand{\BIBdecl}{\relax}
\BIBdecl

\bibitem{Smart_Radio_Environments}
M.~Di~Renzo, A.~Zappone, M.~Debbah, M.-S. Alouini, C.~Yuen, J.~de~Rosny, and S.~Tretyakov, ``Smart radio environments empowered by reconfigurable intelligent surfaces: How it works, state of research, and the road ahead,'' \emph{{IEEE} J. Sel. Areas Commun.}, vol.~38, no.~11, pp. 2450--2525, 2020.

\bibitem{vuyyuru2023finite}
S.~K.~R. Vuyyuru, R.~Valkonen, S.~A. Tretyakov, and D.-H. Kwon, ``Efficient synthesis of passively loaded finite arrays for tunable anomalous reflection,'' \emph{arXiv preprint arXiv:2312.04441}, 2023.

\bibitem{MacroscopicARM2021}
A.~Díaz-Rubio and S.~A. Tretyakov, ``Macroscopic modeling of anomalously reflecting metasurfaces: Angular response and far-field scattering,'' \emph{{IEEE} Trans. Antennas Propag.}, vol.~69, no.~10, pp. 6560--6571, 2021.

\bibitem{Tang2021}
W.~Tang, M.~Z. Chen, X.~Chen, J.~Y. Dai, Y.~Han, M.~Di~Renzo, Y.~Zeng, S.~Jin, Q.~Cheng, and T.~J. Cui, ``Wireless communications with reconfigurable intelligent surface: Path loss modeling and experimental measurement,'' \emph{IEEE Transactions on Wireless Communications}, vol.~20, no.~1, pp. 421--439, 2021.

\bibitem{Tang2022}
W.~Tang, X.~Chen, M.~Z. Chen, J.~Y. Dai, Y.~Han, M.~D. Renzo, S.~Jin, Q.~Cheng, and T.~J. Cui, ``Path loss modeling and measurements for reconfigurable intelligent surfaces in the millimeter-wave frequency band,'' \emph{IEEE Transactions on Communications}, vol.~70, no.~9, pp. 6259--6276, 2022.

\bibitem{Huang2022}
J.~Huang, C.-X. Wang, Y.~Sun, R.~Feng, J.~Huang, B.~Guo, Z.~Zhong, and T.~J. Cui, ``Reconfigurable intelligent surfaces: Channel characterization and modeling,'' \emph{Proceedings of the IEEE}, vol. 110, no.~9, pp. 1290--1311, 2022.

\bibitem{Vittorio2022}
V.~Degli-Esposti, E.~M. Vitucci, M.~D. Renzo, and S.~A. Tretyakov, ``Reradiation and scattering from a reconfigurable intelligent surface: A general macroscopic model,'' \emph{IEEE Transactions on Antennas and Propagation}, vol.~70, no.~10, pp. 8691--8706, 2022.

\bibitem{Vitucci2023}
\BIBentryALTinterwordspacing
E.~M. Vitucci, M.~Fabiani, and V.~Degli-Esposti, ``Use of a realistic ray-based model for the evaluation of indoor rf coverage solutions using reconfigurable intelligent surfaces,'' \emph{Electronics}, vol.~12, no.~5, 2023. [Online]. Available: \url{https://www.mdpi.com/2079-9292/12/5/1173}
\BIBentrySTDinterwordspacing

\bibitem{Vienna5GSLS}
M.~K. Müller, F.~Ademaj, T.~Dittrich, A.~Fastenbauer, B.~R. Elbal, A.~Nabavi, L.~Nagel, S.~Schwarz, and M.~Rupp, ``Flexible multi-node simulation of cellular mobile communications: the {Vienna 5G System Level Simulator},'' \emph{EURASIP Journal on Wireless Communications and Networking}, vol. 2018, no.~1, p.~17, Sep. 2018.

\bibitem{Hao2022}
L.~Hao, A.~Fastenbauer, S.~Schwarz, and M.~Rupp, ``Towards system level simulation of reconfigurable intelligent surfaces,'' in \emph{2022 International Symposium ELMAR}, 2022, pp. 81--84.

\bibitem{Hao2023}
L.~Hao, S.~Schwarz, and M.~Rupp, ``The extended {Vienna} system-level simulator for reconfigurable intelligent surfaces,'' in \emph{{2023 Joint European Conference on Networks and Communications \& 6G Summit (EuCNC/6G Summit)}}, 2023, pp. 1--6.

\bibitem{Sihlbom2022}
B.~Sihlbom, M.~I. Poulakis, and M.~D. Renzo, ``Reconfigurable intelligent surfaces: Performance assessment through a system-level simulator,'' \emph{IEEE Wireless Communications}, pp. 1--10, 2022.

\bibitem{vuyyuru}
S.~K.~R. Vuyyuru, R.~Valkonen, D.-H. Kwon, and S.~A. Tretyakov, ``Efficient anomalous reflector design using array antenna scattering synthesis,'' \emph{{IEEE} Antennas Wireless Propag. Lett.}, vol.~22, no.~7, pp. 1711--1715, 2023.

\bibitem{Sergei2023}
S.~Kosulnikov, F.~S. Cuesta, X.~Wang, and S.~A. Tretyakov, ``Simple link-budget estimation formulas for channels including anomalous reflectors,'' \emph{IEEE Transactions on Antennas and Propagation}, vol.~71, no.~6, pp. 5276--5288, 2023.

\bibitem{balanis2015antenna}
C.~A. Balanis, \emph{Antenna Theory: Analysis and Design}.\hskip 1em plus 0.5em minus 0.4em\relax John wiley \& sons, 2015.

\bibitem{Ellingson2021}
S.~W. Ellingson, ``Path loss in reconfigurable intelligent surface-enabled channels,'' in \emph{2021 IEEE 32nd Annual International Symposium on Personal, Indoor and Mobile Radio Communications (PIMRC)}, 2021, pp. 829--835.

\bibitem{Wu2018}
Q.~Wu and R.~Zhang, ``Intelligent reflecting surface enhanced wireless network via joint active and passive beamforming,'' \emph{IEEE Transactions on Wireless Communications}, vol.~18, no.~11, pp. 5394--5409, 2019.

\bibitem{Rubio2021}
A.~Díaz-Rubio and S.~A. Tretyakov, ``Macroscopic modeling of anomalously reflecting metasurfaces: Angular response and far-field scattering,'' \emph{IEEE Transactions on Antennas and Propagation}, vol.~69, no.~10, pp. 6560--6571, 2021.

\end{thebibliography}
